\documentclass[
  journal=large,
  manuscript= Research Article,
  year=2024
]{cup-journal}

\usepackage[nopatch]{microtype}
\usepackage{booktabs}
\usepackage{amsmath,amssymb,amsfonts}
\usepackage{algorithmic}
\usepackage{comment}
\usepackage{graphicx}
\usepackage{textcomp}
\usepackage[dvipsnames]{xcolor}
\usepackage[hidelinks]{hyperref}
\usepackage{url}
\usepackage{tabularx}
\usepackage{nameref}
\usepackage{wrapfig}
\usepackage[table]{xcolor}
\usepackage[inkscapearea=page]{svg}
\usepackage[backend=biber, style=ieee, sorting=none]{biblatex}
\title{A Novel Idea Generation Tool using a Structured Conversational AI (CAI) System}

\author{B. Sankar}
\affiliation{Department of Mechanical Engineering, Indian Institute of Science (IISc), Bangalore, India - 560012}
\email[B. Sankar]{sankarb@iisc.ac.in}

\author{Dibakar Sen}
\affiliation{Department of Design and Manufacturing (erstwhile CPDM), Indian Institute of Science (IISc), Bangalore, India - 560012}

\keywords{artificial intelligence (AI), generative pretrained transformer (GPT), ideation, large language model (LLM), product design, conversational AI (CAI)} %% First letter not capped

\begin{document}

% \title{A Novel Idea Generation Tool using a Structured Conversational AI (CAI) System\\
% {\footnotesize \textsuperscript{}}
% \thanks{}
% }

% \author{\IEEEauthorblockN{1\textsuperscript{st} B. Sankar}
% \IEEEauthorblockA{\textit{Department of Mechanical Engineering} \\
% \textit{Indian Institute of Science (IISc),}\\
% Bangalore, India - 560012 \\
% sankarb@iisc.ac.in\\ https://orcid.org/0000-0001-5844-6273}
% \and
% \IEEEauthorblockN{2\textsuperscript{nd}Dibakar Sen}
% \IEEEauthorblockA{\textit{Department of Design and Manufacturing (erstwhile CPDM)} \\
% \textit{Indian Institute of Science (IISc),}\\
% Bangalore, India - 560012 \\
% dibakar@iisc.ac.in \\https://orcid.org/0000-0002-4464-3783}
% }

% \maketitle

\begin{abstract}
This paper presents a novel conversational AI-enabled active ideation interface as a creative idea-generation tool to assist novice designers in mitigating the initial latency and ideation bottlenecks that are commonly observed. It is a dynamic, interactive, and contextually responsive approach, actively involving a large language model (LLM) from the domain of natural language processing (NLP) in artificial intelligence (AI) to produce multiple statements of potential ideas for different design problems. Integrating such AI models with ideation creates what we refer to as an \emph{Active Ideation scenario}, which helps foster continuous dialogue-based interaction, context-sensitive conversation, and prolific idea generation. A pilot study was conducted with thirty novice designers to generate ideas for given problems using traditional methods and the new CAI-based interface. The key parameters of fluency, novelty, and variety were used to compare the outcomes qualitatively by a panel of experts. The findings demonstrated the effectiveness of the proposed tool for generating prolific, diverse and novel ideas. The interface was enhanced by incorporating a \emph{prompt-engineered structured dialogue style} for each ideation stage to make it uniform and more convenient for the designers. The resulting responses of such a structured CAI interface were found to be more succinct and aligned towards the subsequent design stage, namely conceptualization. The paper thus established the rich potential of using Generative AI (Gen-AI) for the early ill-structured phase of the creative product design process.
\end{abstract}

% \begin{IEEEkeywords}
% artificial intelligence (AI), generative pretrained transformer (GPT), ideation, large language model (LLM), product design, conversational AI (CAI)
% \end{IEEEkeywords}

%========================================================================

\section{1. Introduction}
\label{sec:introduction}
Creativity and innovation are quintessential during the conceptual design phase for solving problems effectively in an increasingly complex and technology-driven world. Generating many novel ideas quickly is not only valued but also expected from the designers~(\cite{b1, S2023}). However, this creative ideation process can often be challenging for product designers. Novice designers face this even more daunting, especially in the early idea generation phase~(\cite{b2}). This is because the young designers lack experience and suffer from design fixation, mental block and/or cognitive fatigue during ideation~(\cite{b3, bengtsson2019we}). We collectively refer to these cognitive barriers as \emph{Ideation bottlenecks} that hinder their ability to generate novel ideas (Detailed in the following section \nameref{sec:background_work}).

%\subsection{Creativity and Ideation}
Creativity in product design is recognized as generating novel and useful ideas, a concept embraced by design researchers~(\cite{b5}). During ideation, designers are encouraged to engage in expansive thinking while also considering the practical utility of their ideas, which is crucial for addressing real-world challenges~(\cite{b6, b7}). Creative thinking in this context is an interplay between divergent and convergent thinking, where each cycle of convergence can lead to the emergence of new ideas, supporting a step-by-step approach to problem-solving~(\cite{b4}). Ideation is a critical phase in the early stages of product design, where the aim is to generate a diverse array of ideas, navigating through a dynamic "design space" that evolves from abstract concepts to tangible products~(\cite{b2, b8}). It is the activity of generating, developing, and communicating abstract, ambiguous, and imprecise ideas. The activity typically starts with defining the product function and sub-functions, generating ideas for these components, and then integrating them to form a cohesive concept~(\cite{b9}).

\CUPTWOCOL 
%%% END OF FIRST PAGE IN LARGE-LAYOUT NEEDS TO BE MARKED.%%%

%\subsection{Text as a Mode of Description of Ideas}
In design research, the transition from written information to keyword-based online resource retrieval has become pivotal for contemporary designers. It underlines the importance of text in transforming thinking modes and facilitating systematic ideation through the development of a high-level design language~(\cite{b10}). Keywords play a significant role in this process, guiding designers' self-dialogue and information search behaviours, which commonly involve word and image associations that contribute to generating creative design ideas~(\cite{b9, b11}). Moreover, the text is a reflective tool that aids in forming ideas, with designers contemplating relevant words before searching, indicating that thoughtful consideration precedes information retrieval~(\cite{b11, Singh_Tomar_2023}). The ideation process benefits from external stimuli such as language terms, enhancing the creative potential of design solutions in terms of function, behaviour, and structure~(\cite{b12, b14}).

%\subsection{Traditional Ideation Techniques}
Ideation techniques are integral to the design process and provide designers with a pathway to foster creativity. Over time, product design has developed and employed a wide range of ideation techniques. The Table~\ref{tab.ideation_methods} below summarizes the most commonly used traditional ideation technique, their processes, limitations, cognitive principles involved and their expected outcome with respect to their contribution in generating creative ideas~(\cite{b3, b30, b33, b34, b35, b36, b37, b38, b39, b40, b41, b42, b43, b44, b45, tsai2011design, srinivasan2023analogies}).

\begin{table*}[htbp]
\caption{Common Ideation Techniques, Their Limitations, Cognitive Principles and Outcomes}
\begin{center}
\rowcolors{2}{lightgray!80}{lightgray!40}
\begin{tabular}{|m{2cm}|m{3.3cm}|m{3.3cm}|m{3.0cm}|m{3.0cm}|}
\hline
\centering
{\textbf{Technique}}&
\centering
{\textbf{Process}}&
\centering
{\textbf{Limitations}}&
\centering
{\textbf{Cognitive Principle}} &
{\textbf{Outcome}} \\
\hline
Brainstorming	&	Encourages idea generation in groups and innovation through collaboration	&	Limits individual creativity due to fear of judgment, criticism, or peer pressure & Divergent Thinking & Broad Range of Solutions	\\
\hline
Mind mapping	&	Assists in visualizing the brainstorming process but does not actively generate ideas	&	Does not contribute to ideation directly, leaving a creative burden on the designer & Stimulated/Associative Thinking & Visual Organization showing Relations	\\ %Associative Thinking
\hline
Random / Trigger Word	&	Involves using random words as triggers for new ideas	&	Only prompts designer to derive links between word and problem & Stimulated/Associative Thinking & Bunch of Words	\\
\hline
Rapid Sketching & Quickly producing large quantities of visual representations & skill-dependent, random surface-level exploration without purpose & Spatial/Visual Thinking & Series of dissociated scribbles \\ 
\hline
Prototyping	&	Helps explore, test, and visualize different solutions	& Dissociates from the original context and forces fitting of concepts to problems & Spatial/Visual Thinking & Tangible Models representing a phenomenon	\\
\hline
Synectics	&	Comprehensive method using analogies and metaphors to generate innovative ideas	& Requires significant time and effort to facilitate and may not always yield practical ideas & Analogical/Metaphorical Thinking & Equivalent Creative Solutions from Other Domains	\\
\hline
Analogous thinking	&	Helps draw comparisons between unrelated things to create innovative concepts & Relies on the designer's knowledge to find relevant analogies, limiting its applicability. & Analogical/Metaphorical Thinking & Solutions from analogous situations	\\
\hline
Bio-Inspiration	&	Draws inspiration from nature and biological systems for novel and sustainable design solutions	&	Provides a potent source but does not create ideas itself & Analogical/Metaphorical Thinking & Solutions that emulate from Biological Forms, processes and ecosystems	\\
\hline
Idea-Inspire	&	A systematic approach using abstraction, analogies, and principles from nature for design ideation	&	Provides a tool facilitating creativity but does not independently produce creative ideas & Analogical/Metaphorical Thinking & Solutions from a Repository	\\
\hline
SCAMPER	&	Guides designers to think differently by examining a product from various perspectives	& Leads to superficial and incremental ideation rather than truly innovative ideas & Lateral Thinking & A Set of verbs defining actions	\\ %Critical Thinking
\hline
Six Thinking Hats	&	Encourages thinking from different perspectives	&	Guiding methodology that does not facilitate direct ideation & Lateral Thinking & Well-rounded view of problem  from multiple perspectives	\\
\hline
Heuristic Ideation	&	Stimulates creative ideas using heuristics	&	Results in ideas that lack depth or feasibility & Intuitive Thinking & Educated Guesses\\
\hline
Five Whys	&	Primarily used to understand cause-effect relationships	&	Oversimplifies complex issues and miss underlying systemic problems & Intuitive Thinking & Deeper understanding of the problem\\
\hline
Morphological Chart	&	Assists in systematic exploration of new combinations and innovations	&	Becomes complex and overwhelming when dealing with numerous variables and options, potentially hindering idea generation & Systematic/Structured Thinking & Matrix of solutions across multiple dimensions\\
\hline
SWOT analysis	&	Helps identify strengths, weaknesses, opportunities, and threats related to a project	&	Static tool that creates a platform for analysis but does not assist in generating ideas & Systematic/Structured Thinking & Structured overview of strengths, weaknesses, opportunities and threats	\\
\hline
TRIZ & Delivers a systematic approach to solving problems by identifying contradictions and providing solutions & Provides a framework for problem-solving, does not actively contribute to creative ideation & Systematic/Structured Thinking & Inventive Principles\\
\hline
Design thinking	&	A user-centric method to empathize with users, understand their needs, and develop problem-solving ideas	&	Sets the stage for a problem-solving mindset that does not provide a creative spark & Empathic Thinking & User-centred solutions	\\
\hline
\end{tabular}
\label{tab.ideation_methods}
\end{center}
\end{table*}

Each of the ideation techniques listed in Table~\ref{tab.ideation_methods} offers a unique approach to stimulate creativity and waits for innovative ideas to spur into the designer's mind. Exploring these traditional ideation techniques reveals their foundational principles and inherently unveils common limitations prevalent in these techniques as follows. These methodologies furnish designers with procedural structures through defined rules, guidelines and/or procedures for generating ideas, albeit without direct involvement in the ideation process~(\cite{b13}). While these techniques have been widely used, they often lack to provide active support for novice designers. The responsibility for originality and diversity in generating novel ideas essentially rests with the designer. Therefore, these techniques often fail to actively aid and engage novice designers who encounter hurdles during the early stages of idea generation.  These methods lack the dynamism to support designers during their initial ideation. Even though they enable unconventional thinking and provide systematic procedures, all these methods possess limitations, encompassing potential judgment biases, constraints in visual representation, a requirement of a wide knowledge base and the prerequisite of expertise on the part of the designer.

%\subsection{Problems in Existing Ideation Techniques}
The ideation phase of the design process still poses challenges for inexperienced and sometimes even experienced designers. The quantification and formalization of generating ideas are active research areas, with many formal methods yet to be realized as computational algorithms. Existing tools focus on stimulating creativity and providing inspiration, leaving designers with limited options for creating quality ideas~(\cite{b9}). The success of an idea often relies heavily on personal experience and innate ability. Although brainstorming and morphological analysis are widely used, they rely on individual bias and experience~(\cite{b15})—other ideation methodologies, including bio-inspired design and design-by-analogy, provide guidance and inspiration. Traditional ideation approaches are mostly based on design principles and design thinking methodologies, while program-based ideations automate and integrate the advantages of different ideation approaches. On the other hand, data-driven ideation relies exclusively on analysing existing data obtained from design experiments~(\cite{b2}).

Ideally, the ideation phase should result in diverse potential designs because multiple variations increase the likelihood of finding novel and innovative solutions~(\cite{b16}). However, engineers often fixate on specific design options early, limiting the variety of designs. This phenomenon is called Design fixation, where designers adhere blindly to a limited group of ideas, negatively impacting creativity and reducing the diversity and quantity of generated design concepts~(\cite{b12}). Knowledge and experience can contribute to design fixation, as our preconceived notions restrict the design thinking process. Novice designers may struggle to generate diverse concepts and often exhibit this phenomenon early in design. Experience and expertise play a significant role in creative idea generation, and designers progress through different levels of expertise, although this progression is not necessarily linear~(\cite{b3}). Therefore, existing ideation techniques face challenges in providing concise methodologies applicable to designers of various experiences. Current Ideation methods, driven by knowledge and experience, hinder creativity and pose a challenge in guaranteeing novel ideas.  Considering these constraints, a more proactive, engaging, and responsive tool is required to overcome the bottlenecks during ideation.

Given these limitations, there is a need for an ideation tool that offers a more immersive, engaging, and stimulating role in idea generation rather than serving as a static framework. This gap in the traditional ideation framework marks the potential for integrating \emph{artificial intelligence (AI)} with the ideation. Particularly, \emph{natural language processing (NLP)}, a subset of AI, has a set of algorithms called \emph{large language models (LLM)} that have the ability to generate text based on input queries. An AI-driven tool, such as the \emph{proposed Conversational AI (CAI)}, could dynamically and contextually interact with designers, pushing them towards expanded realms of thought and igniting more inventive ideas. The purpose of this paper is not to undermine the value of traditional ideation techniques, as they remain pivotal in design. Instead, it is prudent to acknowledge the emergence and integration of AI models in ideation. It serves as a promising development for redesigning the creativity and innovation landscape in product design, particularly for designers at the inception of their careers.  To provide such a solution, this study delves into the design, development and potential use case of a Conversational AI (CAI) system to facilitate abundant, novel and diverse idea generation for emerging designers. The proposed tool employs a state-of-the-art large language model called as the \emph{Generative Pretrained Transformer (GPT)}, which was customized and used as an Active Ideation Chatbot to engage designers. It acts as an expert omniscient in ideation, enabling a naturalized human-like conversation. This gives the designers a fresh perspective and invokes new thoughts during the creative process. Therefore, it significantly enhances novice designers' abilities to generate novel ideas for solving complex problems. The present research focuses on comparing the efficiency of the CAI-based Ideation tool with traditional methods. The ideas generated by the proposed tool are compared against the commonly used ideation methods, where the responsibility for idea generation lies with the designer. The shortcomings were identified from the comparison, and then the tool was redesigned to provide structured responses based on structured queries. Different examples of such a structured system for some design problems were also given to show the potential of the structured CAI (s-CAI) system. This paper aims to explore the potential of CAI-enabled ideation, which we refer to as \emph{Computer-Generated Ideation}, to help novice designers overcome ideation bottlenecks, unlock their creative abilities, and provide innovative solutions.

%=======================================================================

%\hfill \hrule

\section{2. Background Work}
\label{sec:background_work}

\subsection{2.1. Identification of Ideation Bottlenecks}
\label{sec:ideation_bottlenecks}
\begin{figure}[htbp]
\centerline{\includegraphics[width=1.0\columnwidth]{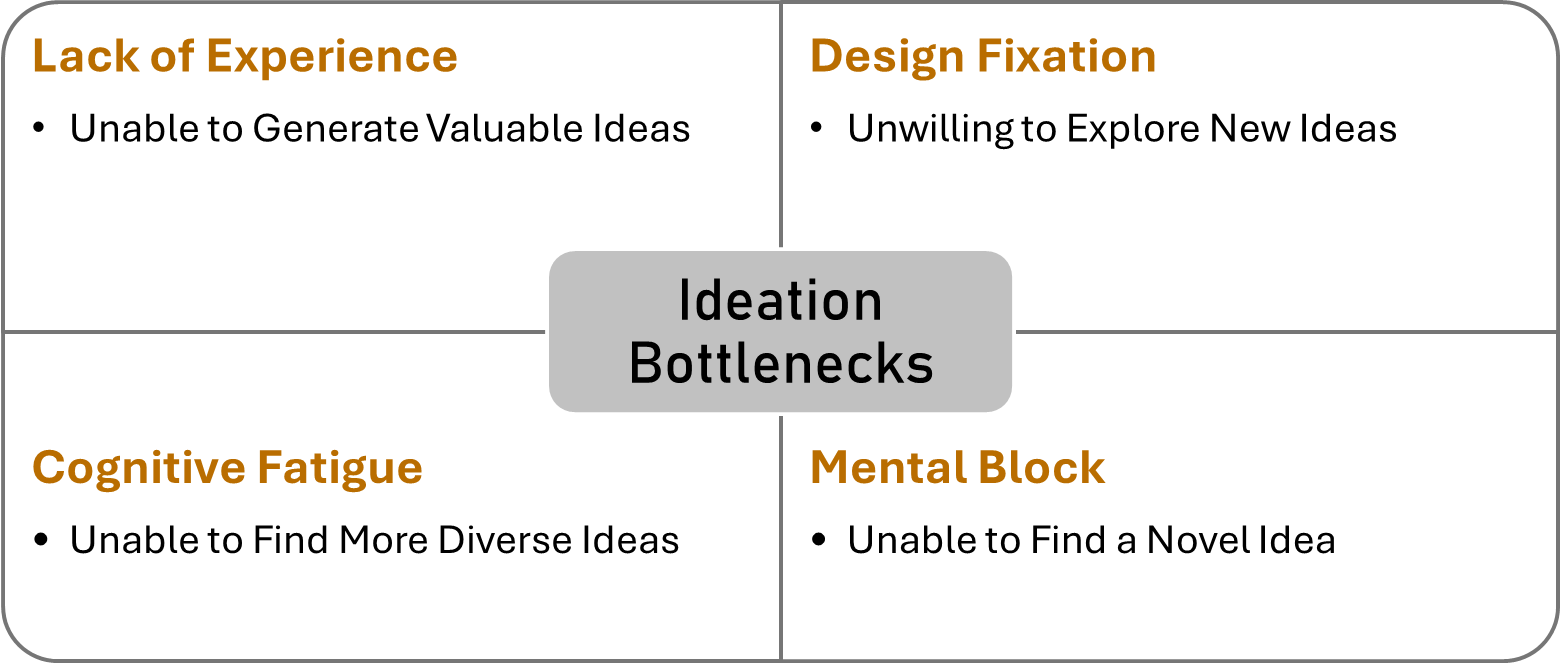}}
\caption{Types of Ideation Bottlenecks}
\label{fig.ideation_bottlenekcs}
\end{figure}

In pursuing innovation through the ideation phase of product design, several impediments can hinder designers' creativity and idea generation. We identify such impediments and categorize them into four primary types, which we term as \emph{Ideation bottlenecks} as summarized in Figure~\ref{fig.ideation_bottlenekcs}. These bottlenecks collectively identified from literature can stifle a designer in creative idea generation. The subsequent explanations of identified bottlenecks are defined by the authors to create a shared understanding within the scope of this study.

%\subsection{Lack of Experience}
\emph{Lack of Experience} is a bottleneck where designers may struggle to generate valuable ideas due to insufficient exposure to varied design challenges or a limited knowledge base. This inexperience can lead to a paucity of ideas or the generation of ideas that lack depth and practical applicability, ultimately impacting the quality and novelty of the design solutions proposed.

%\subsection{Design Fixation}
\emph{Design Fixation} occurs when designers become anchored to their initial ideas. After conceiving a few preliminary concepts, there may be a reluctance or inability to diverge from these nascent solutions. This fixation not only restricts exploration but also inhibits the consideration of alternative and potentially superior ideas, thereby limiting the scope of the design and its potential for innovation.

%\subsection{Cognitive Fatigue}
\emph{Cognitive Fatigue} is a state where the mental exertion of continuous ideation leads to a depletion of cognitive resources. This fatigue can manifest as a reduced ability to generate a breadth of diverse ideas. It can impede the designer’s ability to make connections between disparate ideas, resulting in a homogenization of ideas that fails to push the boundaries of the design solution space.

%\subsection{Mental Blockade}
\emph{Mental Block} represents a bottleneck where designers find themselves unable to conceive novel ideas. This blockage is often characterized by a blank state of mind, where no new ideas emerge, or a cyclic return to previously discarded ideas. Such a block can be due to various factors, including stress, pressure to innovate, or over-saturation with the problem at hand, leading to a paralysis of the creative process.

Each of these bottlenecks presents a significant challenge in the ideation phase, requiring strategies and interventions tailored to address and overcome them. Recognizing these bottlenecks is the first step toward mitigating their effects and fostering a more fluid and dynamic ideation process.

%========================================================================

\subsection{2.2 Classification of Ideation based on their characteristics}
\label{sec:classification_ideation}
In the context of product design, the ideation phase is pivotal yet often encounters many cognitive and practical barriers~(\cite{b31}). Regarding creativity and ideation, various definitions exist, as discussed in the introduction. Still, for the present context, we define a \emph{creative idea as "an assertive statement that describes 'what' part of the solution and needs to be both novel and diverse"}. Assessing the degree of novelty in an idea involves measuring how much it deviates from existing solutions and emphasizes the significance of originality, whereas the diversity of ideas refers to how far they are situated in the design space from one another. In this section, the authors identify certain characteristics prevalent among the traditional ideation techniques, characterized by a reliance on self-established thought patterns and conventional thought-provoking techniques. We introduce characteristics of a new ideation methodology that encourages dynamic engagement and the proactive pursuit of novel solutions using technology. Examining these characteristics offered a nuanced understanding of the ideation tools and helped formulate the proposed ideation method as a novel idea generation tool.

\subsection{2.2.1. Solo Ideation}
\label{sec:solo_ideation}
Novel ideas generated during ideation form the foundation for innovative and impactful products. However, many existing ideation techniques require the designers to work alone. This type of solo ideation is what we term \emph{Passive Ideation}. Passive ideation is defined as one that enables only one-way communication, putting the burden on the designer's mind to generate ideas, thereby limiting the potential of designers. The following are some of the limitations of passive ideation methods.

%\subsubsection{3.1.1. Lack of Engagement}
\emph{Lack of Engagement}: Traditional ideation methods often lack the level of engagement necessary to inspire and ignite the creative process. While valuable, brainstorming or mind-mapping techniques can become monotonous. This passive approach may hinder designers from fully exploring unconventional or breakthrough ideas, restricting their creativity.

%\subsubsection{3.1.2. Limited Collaboration}
\emph{Limited Collaboration}: Many traditional ideation methods are primarily based on personal thought discovery, restricting the power of collaboration. While individual thinking is essential, harnessing the collective intelligence of a diverse group can lead to more innovative concepts. Passive methods may not facilitate effective collaboration, resulting in missed opportunities for cross-pollination of ideas and fresh perspectives. Collaboration also requires discussion with experts and peers. Humans, with their diverse personalities, may hinder the generation of novel ideas as they can be biased towards their ideas.

%\subsubsection{3.1.3. Slow Iteration}
\emph{Slow Iteration}: Passive ideation methods often require significant time investment, slowing the iterative design process. Waiting for individual ideas to be shared, analysed, and consolidated can delay progress and hinder the exploration of alternative concepts. Designers need a more agile, active approach that allows for rapid idea generation, testing, and refinement.

\subsection{2.2.2 Collaborative Ideation}
\label{sec:collaboration_ideation}
Collaborative ideation is characterised by a natural, dynamic, unbiased alliance between a designer and the ideation method, facilitating creativity and critical thinking. This continuous conversation-driven approach serves as a solution to address the limitations of passive ideation methods. This type of Ideation is what we refer to as \emph{Active Ideation}. By embracing active ideation, designers can overcome the constraints of passive ideation techniques, encouraging diverse perspectives and interactive engagement throughout the ideation process.

%\subsubsection{3.2.1. Enhanced Engagement}
\emph{Enhanced Engagement}: A higher level of engagement from the medium characterizes active ideation. Incorporating textual stimuli, dynamic prompts, and immersive experiences, designers are encouraged to think more creatively, break free from conventional thinking, and explore novel solutions to design challenges.

%\subsubsection{3.2.2. Dynamic Idea Exploration}
\emph{Dynamic Idea Exploration}: An active ideation medium encourages designers to think beyond traditional boundaries and experiment with various ideas through multiple quick revisions. By providing the ability to have natural conversational dialogues with an omniscient, designers can explore different solutions and generate multiple design options that would not have been possible.

%\subsubsection{3.2.3. Data-Driven Insights}
\emph{Data-driven Insights}: Active ideation can collect and analyse conversational data during the ideation process, providing designers with valuable insights into the effectiveness and feasibility of different concepts. By leveraging real-time answers, designers can make informed decisions, prioritise ideas, and identify emerging patterns, leading to more diverse and detailed ideas.

%========================================================================

\section{3. Potential of AI for Ideation}
\label{sec:potential_of_ai_for_ideation}
Matching user requirements with novel solutions is a significant challenge due to the multidisciplinary knowledge required by the designer to create potential ideas. Generating many ideas and identifying the most valuable ones is beneficial to developing novel solutions in complex systems and competitive markets~(\cite{b18}). The ideation process has traditionally been a human-centric task, relying on the cognitive abilities of the designer. This approach has significant challenges due to its dependence on designers' expertise and the above bottlenecks. With the use of computers in different phases of design, there has been a progressive shift towards integrating computer-aided tools into the creative ideation workflow as well.

\subsection{3.1. Computer-Aided Ideation}
\label{sec:computer_aided_ideation}
Computer-aided ideation tools have existed for some time, offering a framework and environment where designers can explore, manipulate, and visualize ideas. These tools often incorporate databases of knowledge, templates for brainstorming, and mechanisms for capturing and categorizing ideas~(\cite{b17}). Some of these tools, such as iDea of~\cite{Ekstrmer2019AFS}, Digital Brainstorming of~\cite{Maaravi2021-ic, bryant2005a, Siegle2020-lv}, Idea-Inspire of~\cite{b45}, Bio-Inspire of~\cite{b40}, FuncSION of~\cite{pal2014a}, Co-storm of~\cite{zhang2019acostorm}, PANDA of~\cite{roderman1993a}, Web-enabled ideation of~\cite{Beretta2018vo} etc., have provided significant assistance in the ideation process, helping designers to organize thoughts, inspire creativity, and document the process. They augment the designer's natural ideation capabilities by providing a digital space for exploration and documentation. However, the creative spark and the inception of novel ideas still originate from the human intellect, with the computer acting as a repository and mediator rather than a generator of ideas.

\subsection{3.2. CAI for Collaboration with Computers}
\label{sec:cai_for_collaboration_with_computers}
Let us imagine a situation where computers take a more \emph{active role in idea generation} rather than being just a facilitator in the ideation process as in computer-aided ideation. The computer would generate ideas, suggest alternatives, and even challenge the designer's assumptions, and the designer would play the role of a \emph{curator} of what the computer proposed. That is, the designer's expertise is utilized to evaluate the ideas and select the potential ones. This symbiotic relationship can be viewed as a Human-Computer Collaboration that capitalizes on the strengths of both computers and humans, combining raw computational power and intuitive judgment.

%\subsubsection{3.2.1. LLM in Ideation}
Recently, large language models (LLMs) from the Generative AI domain have gained prominence for their ability to generate diverse forms of original text~(\cite{b19}). It is known that conversation AI (CAI) systems can be tailored to align with human values, as they are designed to emulate intelligent human agents~(\cite{b19}). A salient feature of popular CAI systems is their focus on natural language interaction through chat-based dialogues, resembling human-to-human interactions~(\cite{b19}). In addition, digital big data has opened up novel avenues for such CAI by harnessing the potential of large datasets as knowledge bases. The authors feel that this ability of modern CAI systems to meaningfully respond to vague and grammatically inaccurate queries while exploiting the big-data resource makes them akin to the behaviour of human experts and can, therefore, be exploited for creative problem-solving.

\subsection{3.3. Computer-Generated Ideation}
\label{sec:computer_generated_ideation}
\begin{figure}[htbp]
\centerline{\includegraphics[width=0.9\columnwidth]{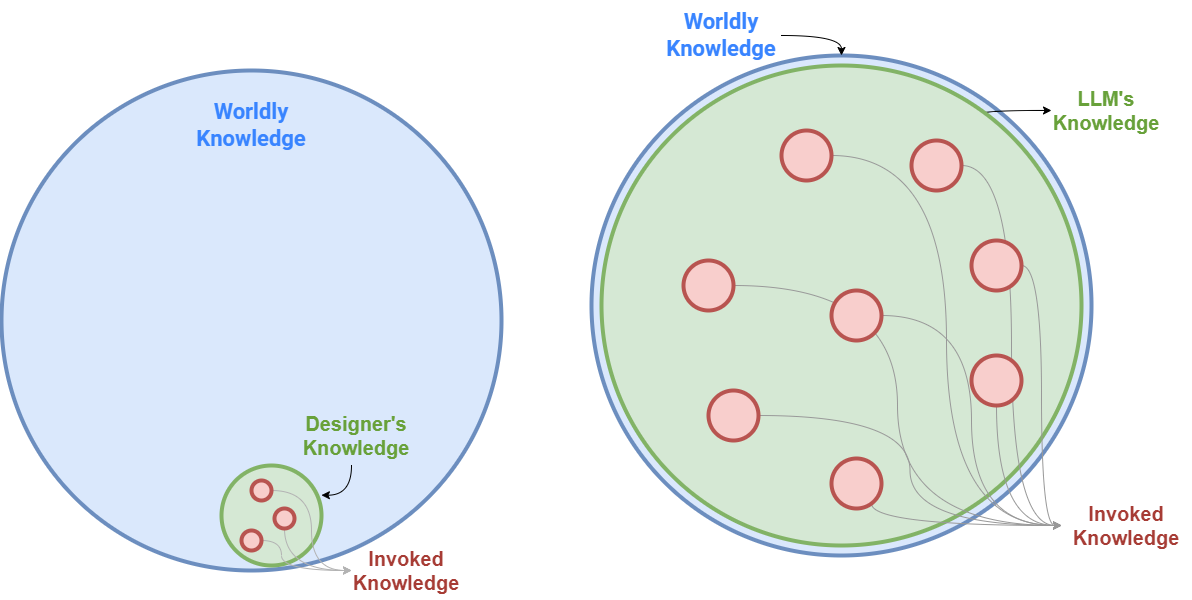}}
\caption{Abstract Representation of Designer's Mind (Left) vs LLM's Mind (Right)}
\label{fig.knowledge_base}
\end{figure}

Ideation in design can be conceptualized as a cognitive exploration that seeks to bridge the gap between the problem and solution spaces. Traditionally, this process relies on the designer's ability to access and leverage their knowledge and experience. Only a fraction of the designer's knowledge, which is a small subset of the world's collective knowledge as shown in Figure~\ref{fig.knowledge_base}, gets invoked at the crucial moment for idea generation. This inherent limitation can significantly constrain a designer's capacity to generate novel, high-quality ideas within a given time. Moreover, the spontaneous invocation of pertinent knowledge is not within the designer's volitional control, which can lead to cognitive stress and fatigue. This unpredictable nature of retrieval of a designer's own knowledge underscores the challenges faced in conventional ideation techniques. Therefore, we believe that timely access and harvesting of principles from a large repository of knowledge is critical for the formulation of creative solutions for difficult practical problems.

% CAI to CDI
Computer-generated ideation using CAI includes the application of LLMs, which can interact with designers through natural language processing (NLP). These LLM models can understand context, generate coherent ideas, assist in the iterative refinement of ideas, and provide a rationale for their suggestions. The LLMs, such as conversational AI, possess two critical attributes that make them invaluable to the ideation process. First, they have \emph{access to a vast repository of knowledge} far exceeding the scope of any individual designer's memory as shown in Figure~~\ref{fig.knowledge_base}. Second, given appropriate \emph{inputs (Prompts)}, they can rapidly generate coherent and contextually relevant sentences as \emph{outputs (Responses)}. The morphology of these sentences can resemble an idea. Thus,  although LLMs lack intrinsic cognition and the ability to discriminate, their apparently intelligent responses can potentially be exploited fruitfully. This can be done by effectively invoking the relevant knowledge from their extensive database. By integrating such advanced computational technologies as active ideation partners, this CAI-based ideation promises a collaborative effort between human designers and machines. The designer's responsibility shifts towards critically analysing and selecting the best ideas, harnessing the full potential of computational power for the generation of novel ideas.

%========================================================================

\section{4. Ideation through CAI} \label{sec:ideation_cai}
There are two interesting features that make the CAI systems appear to be intelligent: (a) the ability to generate intellectually acceptable write-ups on a given topic and (b) the ability to generate responses to subsequent queries that build upon the previous interactions. This makes the interaction a coherent conversation on a given topic. Thus, if feature-(a) is a description of an idea, then feature-(b) could be constructed as an elaboration and clarification of that idea. In the following paragraphs, we elaborate on how these features of one such CAI system, GPT, can be customized and utilized for conversational design ideation. The concepts so developed are applicable to other CAI systems as well.

\subsection{4.1. GPT – A Conversational AI}
\label{gpt_a_conversational_ai}
The Generative Pre-Trained Transformer (GPT) is a state-of-the-art large language model (LLM) developed by OpenAI. As a natural language model, it has been trained to predict the next word in each piece of text, enabling it to generate coherent and contextually relevant sentences~(\cite{b20}). The GPT model by OpenAI has undergone a complex training regime that involves three core stages: unsupervised pre-training (USPT), supervised fine-tuning (SFT), and reinforcement learning from human feedback (RLHF). These stages collectively train GPT to produce human-like text while adhering to safety and ethical standards, balancing raw performance and controlled behaviour~(\cite{b21}). Due to its vast size and extensive training, the model can answer questions, write essays, summarise texts, and even translate languages - essentially any task that involves predicting the next word in a sentence~(\cite{b22}).

\subsection{4.2. Characteristics of GPT as a potential tool in Ideation}
\label{sec:characteristics_gpt_potential_ideation_tool}
GPT models are pre-trained on vast text corpora, enabling them to generate coherent and contextually relevant text responses based on input prompts. We feel that if the input prompt solicits a solution to the problem, GPT would generate a text that could be a practical idea to solve the problem. This characteristic makes GPT a potential asset during the conceptual phase of product design. By engaging in a dialogue with GPT, designers can articulate their design challenges, ask questions, and receive prompt and personalised responses similar to brainstorming with multiple people, but instead with an expert omniscient. Furthermore, its capacity to process and synthesize information from various domains can help cross-pollinate ideas, thereby fostering innovation. The iterative interaction with a GPT can help novice designers overcome ideation bottlenecks by presenting a flow of ideas that can be refined and expanded upon, ensuring a dynamic and fluid creative process.

For instance, a GPT model might link a problem in ergonomic furniture design with insights from bio-mechanics and psychology, fields that may not typically be associated but can provide a deeper understanding of user interaction and comfort. By establishing such connections, designers are empowered to use knowledge more effectively, applying it in contextually appropriate ways to address the problem. The GPT's role in this process is to act as a cognitive enhancer, expanding the designer's ability to think laterally and draw upon a wider array of interdisciplinary insights, crucial for innovation and developing holistic design solutions.

\subsection{4.3. Potential benefits of using GPT for Ideation}
\label{sec:potential_benefits_gpt_ideation}
The importance of making connections between knowledge in the design solution space cannot be overstated~(\cite{Goncher2009}). Design is inherently an integrative process, requiring the synthesis of various types of knowledge to create solutions that are not only innovative but also practical and feasible~(\cite{b32}). A GPT model can facilitate this synthesis by identifying patterns and relationships within the data they have learned that might not be immediately apparent to human designers. This ability to make unexpected connections can lead to breakthrough ideas and creative leaps in the design process. GPT models are adept at connecting disparate knowledge to specific problems, a key function during the ideation phase of product design. By drawing from a comprehensive database of information, GPT can bridge the gap between abstract concepts and concrete design challenges. When a designer inputs user needs or a design brief into a GPT-powered tool, the AI model can analyze the text, identify key themes and requirements, and then scan its vast repository of learned data to suggest relevant ideas, analogies, and concepts. This ability of GPT to understand and generate diverse linguistic structures allows it to serve as an ideation partner that can offer novel perspectives and solutions that might not be immediately obvious to human designers. The following are the three important benefits of using a CAI such as GPT for ideation:

\subsubsection{4.3.1. Inspiration and Knowledge Expansion}
GPT can provide designers with a wealth of information, offering insights from various industries, scientific and design principles, emerging trends and latest technologies. It acts as a virtual collaborator, inspiring designers by presenting alternative perspectives and facilitating cross-pollination of ideas.

\subsubsection{4.3.2. Rapid Iteration and Feedback}
Designers can quickly iterate their ideas by asking for feedback from GPT. The model's ability to understand and respond to nuanced queries allows for immediate evaluation and refinement of design concepts. This accelerates the ideation process and enables designers to explore broader possibilities.

\subsubsection{4.3.3. Contextual Guidance}
GPT can guide designers by asking relevant questions, challenging assumptions, and offering suggestions. By providing contextual guidance, the model helps designers think critically, consider multiple factors, and refine their design ideas with a holistic perspective.

\subsection{4.4. Design of CAI-based Ideation Interface}
\label{sec:ideation_interface}
\begin{figure*}[htbp]
\centerline{\includegraphics[width=1.0\textwidth]{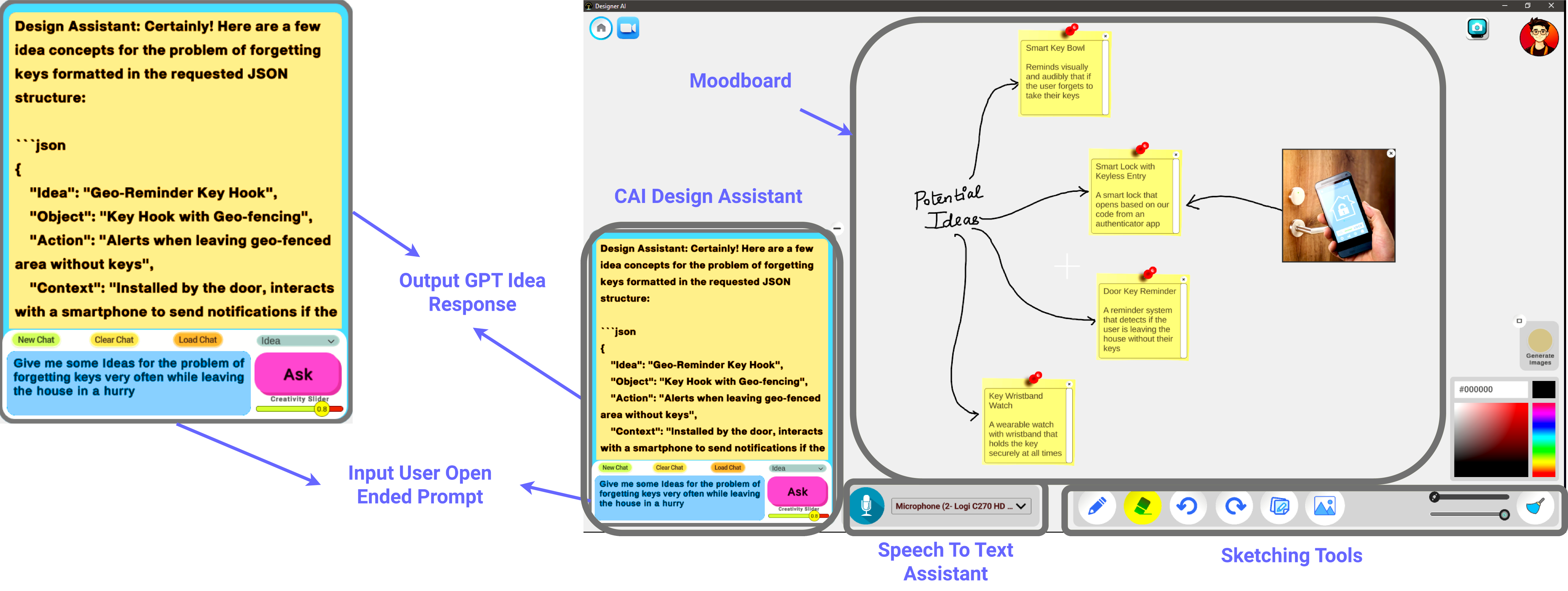}}
\caption{Interface of the Conversational AI-based Active Ideation Tool}
\label{fig.interface}
\end{figure*}

An active ideation interface was designed and developed using a conversational AI system known as a Generative Pretrained Transformer (GPT)\footnote{We used GPT-4 in our implementation.} embedded over an interactive moodboard, as shown in Figure~\ref{fig.interface}. GPT provides the backbone for natural language interaction, allowing it to respond and generate creative idea statements based on user input. The mood board provides a means for rapidly putting down those ideas. Thus, the interface provides the designers with a conversational and intuitive platform where GPT drives idea generation.

\subsubsection{4.4.1. Fine-tuning GPT as an Expert Designer}
The GPT model was fine-tuned to serve as an expert designer for active ideation, which involves initializing the model with a vast and diverse corpus of text data from various relevant online sources, viz., product design-related documents, books, and articles. This exposure to a wide range of design concepts, methodologies, paradigms, and approaches allows the model to develop a broad understanding of the design field. This process can also enable the model to learn the domain-specific language, creative design ideation, innovative problem-solving skills, and various design thinking strategies applied by expert designers. By doing so, the fine-tuning phase essentially imparts the language model with the essence of design expertise, allowing it to generate insightful, context-specific, and creative ideas in the ideation process\footnote{Any Mention of GPT from here on refers to this custom fine-tuned GPT model}.

\subsubsection{4.4.2. Multi-session Ideation through Contextual Understanding}
A contextual buffer memory was created and used in context continuation tasks, such as design ideation sessions. This enables sustained engagement in a context-aware conversation for a longer duration. This method of learning from an earlier conversation to keep the context of a current conversation is termed \emph{Contextual Understanding}. The contextual buffer memory is designed to store the earlier conversations in a standard JSON file. In the case of an ideation session, if a designer prefers to carry out ideation in multiple sessions/sittings, the contextual buffer memory helps GPT to preserve the context of previous sessions and proceed from where it stopped. Thus the insights or ideas from previous sessions are not lost, and the creative thinking flow remains consistent.

\subsubsection{4.4.3. Nudging GPT for Out-of-the-Box Ideas}
The GPT allows setting a parameter called "temperature" to adjust the randomness of its responses. This randomness, interpreted as an index of novelty, can be used to adjust the creativity of the generated ideas. A higher temperature setting leads to more diverse and imaginative responses, encouraging the exploration of unconventional ideas and stimulating creative thinking. On the other hand, a lower temperature value produces more deterministic and focused responses, aligning closely with known patterns and preferences.

%========================================================================

\section{5. Evaluating Effectiveness of CAI-based ideation}
\label{sec:evaluating_effectiveness_gpt_based_ideation}
The potential of CAI-based ideation envisaged above needs to be empirically validated. Towards this end, the following two research questions are explored through practical design sessions in which graduate design students participated. This study uses the GPT model to create a design chatbot and a moodboard as an ideation interface using Unity and C\# programming. The GPT chatbot engages with the designer in a naturalized conversation using text or voice interaction modality to generate ideas as responses for the input problem statements.

\textit{RQ1. Is there a benefit to using our Conversational AI-embedded ideation over conventional ideation techniques?}

Diverse conventional ideation methods are available to help designers provide relevant information and inspiration to generate ideas relevant to the problem at hand. However, due to their rule-ridden passive nature, these traditional methods impose the burden of generation on the designers. On the contrary, CAI-driven ideation takes an active role where the computer takes up the generation, and the designer's role is in evaluating and selecting potential ideas.

\textit{Methodology for RQ1:} We propose to use GPT to create a design chatbot adjacent to a moodboard forming an ideation interface using Unity and C\# programming. Ideas for solving a given problem are sought from the designer, once through the conventional ideation techniques and once using the CAI-based ideation interface designed by us in a time-restricted format. At the end of the experiment, all the ideas are assessed for their novelty and fluency. We then check if there are any significant differences in these parameters in the two modalities of ideation.

\textit{RQ2. Does Conversational AI-based ideation help novice designers overcome ideation bottlenecks?}
Novice designers transitioning from academia to the professional realm primarily suffer from design fixation and mental blocks when faced with difficult real-world problems that require innovative solutions. These bottlenecks pose a barrier for the designer, confining them to a small solution space. Such prolonged barriers can lead to cognitive fatigue when the designers withdraw from their ideation endeavours. Therefore, there is a need to support novice designers in overcoming their mental blocks by providing the ability to generate diverse ideas from multiple solution spaces.

\textit{Methodology for RQ2}: The temperature parameter available in GPT to control the randomness of the responses in terms of their connection to the problem at hand. The designer adjusts this parameter while using the GPT design chatbot and acts on the response by framing elementary ideas in the moodboard provided adjacently. The designer sets a higher temperature setting to get potential responses from unconventional knowledge domains, pending further clarification and elaboration to form the implicit connection. On the contrary, using a lower temperature setting, the response generated itself is likely to be an idea, thereby explicitly showing a connection to the problem without the need for much clarification. The sets of ideas jotted down by the designer are then assessed for their variety.

The aforementioned Research Questions (RQ1 \& RQ2) are answered by formulating three hypotheses (H1, H2 and H3), which are validated by conducting a pilot study as detailed in the following section~(\nameref{sec:pilot_study}).

%========================================================================

\section{6. Pilot Study}
\label{sec:pilot_study}
\begin{figure}[htbp]
\centerline{\includegraphics[width=1.0\columnwidth]{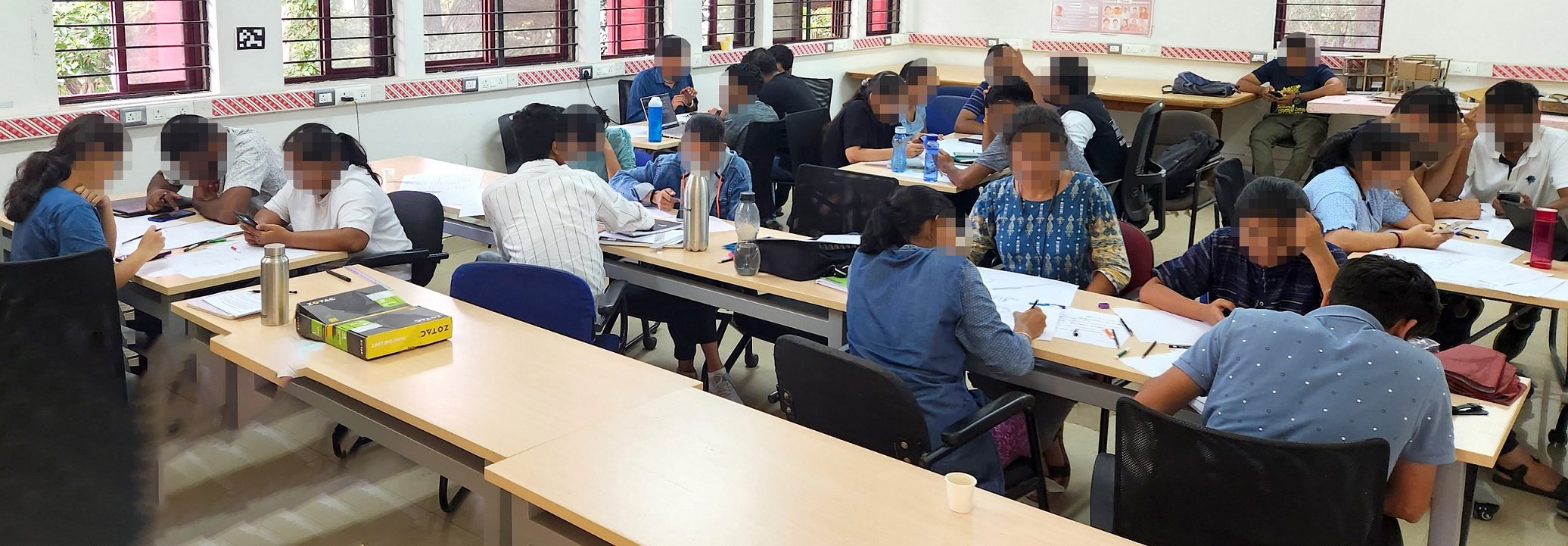}}
\caption{Participants Performing the Study}
\label{fig.participants}
\end{figure}

Since the present study investigates the potential benefit of the proposed CAI-based ideation interface for novice designers, thirty postgraduate product design students were recruited (Figure~\ref{fig.participants}). The participants were randomly divided into six groups, each assigned a specific design problem statement. Before embarking on the design task, all participants received comprehensive training in utilising various conventional ideation methods, and a practice session was conducted to ensure their familiarity with the techniques. Ethical considerations were upheld, and proper consent was obtained from all participants for using the study's outcomes in further research endeavours. The study comprised two distinct parts, Part A and B, described below.

\subsection{6.1. Comparative Ideation Study}
The comparative ideation study is conducted to assess the effectiveness of idea generation when comparing human designers with Conversational AI (CAI). The following sections provide an in-depth look at the study's methodology.

\subsubsection{6.1.1. Performance Metrics}
The assessment of idea generation effectiveness encompasses various commonly employed metrics, including novelty, variety, quality, and quantity~(\cite{b23}). Generating a higher quantity of ideas can lead to the emergence of higher-quality design concepts~(\cite{b24}). As reported in \cite{b23}, Novelty pertains to the degree of uniqueness of an idea, while variety encompasses the extent of exploration within the solution space. Novelty ($\eta$), therefore, signifies the divergence of an idea from existing ones, while variety ($\upsilon$) encapsulates the richness of diversity among ideas. Fluency ($\Gamma$) denotes the quantity of ideas generated within a specified period. The assessment of ideas is essential, as not all generated ideas progress beyond the ideation phase into further development.

\subsubsection{6.1.2. Hypothesis Formulation}
\vskip10pt

\noindent \emph{H1: CAI produces meaningful ideas.}

During ideation, the main task is proposing methods (ideas) that potentially solve a given problem.  Ideas can be communicated through diverse modalities. In the present context, ideas are presented textually. It is known that CAI is capable of generating meaningful textual content. This hypothesis asserts that, with suitable prompts, the GPT-generated statements would constitute meaningful ideas when presented to a designer.\vskip10pt

\noindent \emph{H2: CAI promotes prolific novel ideas.}

In a collaborative ideation scenario, the designers continuously engage in a conversation focused on a given problem. The output of such engagement is to generate multiple solutions using their collective knowledge. The level of involvement among designers during this scenario is characterised by the uninterrupted exchange of ideas. The effectiveness of such a collaboration can be assessed by counting the number of quality ideas generated. This hypothesis asserts that the set of ideas generated through CAI would have higher fluency and novelty than those in an unaided exercise.\vskip10pt

\noindent \emph{H3: CAI promotes diverse ideas.}

Past experiences, events, and presented stimuli help the designer formulate creative ideas to solve a problem. The diversity of the ideas in a set refers to how different the ideas are from each other. High diversity is generally preferred as there is a higher potential for an out-of-the-box idea~(\cite{S2023}). This hypothesis asserts that the diversity of the set of ideas generated through CAI will be higher than the unaided exercise thereby establishing its potential to alleviate ideation bottlenecks.

\subsubsection{6.1.3. Study Design}
A protocol was designed to conduct the study in two phases for each participating designer.

Part A: Ideas were generated by the designers for an unseen design task using traditional ideation techniques such as Brainstorming, Random Words, SCAMPER, Synectics, and Analogous Thinking. The allocated time for this exercise is 20 minutes. Here, there is no usage of a computer or any other digital gadget. The designers had to use their own knowledge resources.

Part B: Participants were instructed to use the CAI-based ideation tool (As discussed in Section ~\nameref{sec:ideation_cai}) for idea generation and use their judgement to shortlist the potential ideas. The allocated time for this exercise is 20 minutes. A training and practice session was held for each participant to familiarise him/her with the interface.

Each group comprising of five designers was randomly allocated one of the following design problem statements(PS).
\begin{itemize}
    \item PS1: Product for segregation as a means for effective waste management
    \item PS2: Product for footwear disinfection and cleaning for improved hygiene and safety
    \item PS3: Product for enhancing household dish cleaning efficiency and sustainability
    \item PS4: Product for enhancing comfort and efficiency for prolonged standing in queues
    \item PS5: Product for bird-feeding for fostering mental well-being of elderly individuals at Home 
    \item PS6: Product for convenient umbrella drying and storage on travel
\end{itemize}
Each of the above problems was elaborated through a slideshow that covered the background, user needs, challenges and requirements so that the designers could readily start on the ideation stage of designing.

\subsubsection{6.1.4. Study Setting}
The ideation exercise was conducted wherein each group was assigned one conventional ideation technique during part-A of the study to generate ideas for the randomly assigned problems. Each group was instructed to create an unordered list of idea statements, contributing as many ideas as they saw fit without any constraint on quantity. It was essential to note that the internal dynamics of the groups, such as the influence of dominant personalities or the collaborative synergy, were not monitored during this exercise. The goal was to purely capture the raw output of ideas generated through traditional and CAI-enabled ideation. The exercise was organized in a way that allowed for a natural flow of creativity and innovation without posing any restrictions to the designer.

The behaviour of groups during both parts A and B of the exercise revealed interesting patterns. Part A focused on the generation of ideas by each designer using the ideation technique assigned to that group. Part B centred around the collective refinement and curation of the ideas generated by GPT as a group. Despite the introduction of CAI in the mix, recognising and selecting potential ideas remained a distinctly human task, underscoring the collaborative nature of the ideation process between CAI and designers. The outcome of this ideation exercise as a part of the pilot study was a curated set of ideas jotted down as textual statements where the CAI tool was used in augmenting the ideation process with a breadth of responses as shown in Table~\ref{tab.set_of_ideas}. The final judgment on the potential of such responses was left to the decision based on the designer's expertise.

\begin{table*}[h!t!bp]
    \centering
    \caption{Set of Idea Statements Generated during Part A and Part B of the study}
    \label{tab.set_of_ideas}
    \rowcolors{2}{lightgray!80}{lightgray!40}
    \begin{tabular}{|m{1.5cm}|m{7.0cm}|m{7.0cm}|}
    \hline
        \centering
       \textbf{Problem Statements (PS)} & 
       \centering
       \textbf{Ideas Generated by Designers using Traditional Ideation Methods (Part-A)} & 
       \textbf{Ideas Generated by GPT in CAI-enabled Ideation (Part-B)}\\
    \hline
        PS1 
        & 
        DI-1.1. Introduce a foot pedal system for bin lids, enabling users to easily open them without using their hands, thus maintaining hygiene.\newline
        DI-1.2. Design bins in distinct colours to instantly signal the type of waste they're for, simplifying the segregation process for users.\newline
        \dots
        & 
        CI-1.1. Stackable Type Bins: Designing stackable bins for efficient storage when not in use is ideal for homes or spaces with limited storage capacity. This design promotes easy access and ensures that bins are readily available when needed.\newline
        CI-1.2. Incentivized Waste Segregation: Implementing an incentive-based system where users receive rewards or benefits for proper waste segregation and disposal. This encourages active participation and responsible waste management.\newline
        \dots\\
    \hline
        PS2 
        & 
        DI-2.1. Design an automated scrubber with motorized bristles to deep-clean shoes, effortlessly removing dirt.\newline
        DI-2.2. Develop a cleansing foam that can be applied to shoes for a quick clean without the need for water.\newline
        \dots
        & 
        CI-2.1. A Mat with Built-in Bristles: Designing entryway mats with built-in bristles or brushes that users can walk across. The bristles would scrape and clean shoe soles, removing dirt and debris before entering a clean environment.\newline
        CI-2.2. Liquid Jet Spray Chamber: Developing footwear cleaning chambers equipped with liquid jet spray mechanisms that shoot cleaning solutions onto shoes. Users would enter the chamber, and the liquid jets thoroughly cleaned and disinfected their footwear.\newline
        \dots\\
    \hline
        PS3 
        & 
        DI-3.1. Install an adjustable water sprinkler over the sink for a hands-free rinse-off of food particles from dishes.\newline
        DI-3.2. Craft a wearable scrubber that allows for hands-free dishwashing.\newline
        \dots
        & 
        CI-3.1. Dishwashing Sink with Rotating Brushes and Water Stream Jet: Integrating rotating brushes and water stream jets within the sink basin facilitates efficient manual dishwashing. Users can scrub dishes while the sink's features assist in cleaning.\newline
        CI-3.2. Dishwashing Gloves: Developing dishwashing gloves with built-in scrubbing surfaces on the palms and fingers allows users to scrub dishes without the need for separate scrubbing tools.\newline
        \dots\\
    \hline
        PS4 
        & 
        DI-4.1. Develop a foldable stool that can be easily carried and deployed for temporary relief during prolonged waits in Queues.\newline
        DI-4.2. Create cushioned shoes designed to reduce fatigue from standing for long periods.\newline
        \dots
        &
        CI-4.1. Portable Ergonomic Support Device: Design a compact portable device that users can carry with them and set up quickly on the ground and lean against for ergonomic support. These devices would have adjustable features to cater to users of different heights and body types.\newline
        CI-4.2. Queueing Cushion: Developing cushions designed specifically for queueing featuring ergonomic shapes and materials that provide comfortable seating options for individuals waiting in line with Inserts Straps.\newline
        \dots\\
    \hline
        PS5 
        &
        DI-5.1. Design a bird feeder with an integrated voice assistant that recognizes and narrates bird species, promoting mental stimulation and learning.\newline
        DI-5.2. Develop an automated feeding bowl that ensures a consistent supply of bird food, minimizing maintenance while maximizing bird-watching opportunities.\newline
        \dots
        &
        CI-5.1. Window Mounted Feeder with One-way Mirror: Designing a bird feeder that mounts to windows with a one-way mirror. Users can observe birds up close without disturbing them, enhancing the sense of connection with nature.\newline
        CI-5.2. Automatic Gravity-Enabled Feeder: Creating an automatic bird feeder that uses gravity to dispense feed as needed, reducing manual refilling efforts and ensuring a consistent food supply.\newline
        \dots\\
    \hline
        PS6 
        &
        DI-6.1. Develop a waterproof casing for umbrellas that prevents drips in entryways after coming indoors from the rain.\newline
        DI-6.2. Introduce a mechanical twister that wrings out the excess water from an umbrella.\newline
        \dots
        &
        CI-6.1. Use and Throw Canopy: Designing disposable umbrella canopies made from eco-friendly materials. Users can replace the canopy after use, reducing the need for drying and storage and minimizing environmental impact.\newline
        CI-6.2. Electrostatic Dryer: Designing umbrellas with an electrostatic drying mechanism that is battery activated and repels water from the canopy when activated, ensuring a dry umbrella before coming indoor.\newline
        \dots\\
    \hline
    \end{tabular}
\end{table*}

\subsubsection{6.1.5. Assessment Method}
As per the popularly used Consensual Assessment Technique (CAT)~(\cite{b1}) for idea evaluation, the end-users assessment and the expert evaluations are both equally important. The ideas generated during the convention and CAI-based ideations from each group were collected, and a Google form was created.  Thus, to evaluate the design ideas generated in both Part A and Part B of the study, they were submitted for assessment to design researchers affiliated with the Department of Design in renowned institutions in India, such as IIT Delhi, IIT Guwahati, IIT Bombay, and IISc Bangalore. A total of 80 responses were collected. The subsequent section will present a comprehensive analysis of these responses. The respondents served dual roles as expert evaluators and end users of the presented solutions. 
%The data collected during the study is accessible using the link provided here \href{https://drive.google.com/drive/folders/1IW4ONgTy4uUFDPxZvWJm2KTH82LgePP-?usp=sharing}{google drive}.

%--------------------------------------------------------------------------------------

\subsection{6.2. Results and Outcome}
The pilot study undertaken in this research was meticulously designed to validate Hypotheses H1, H2, and H3, each positing a different aspect of the effectiveness and utility of Conversational AI (CAI) in the ideation process. The results obtained from this study are critical in determining the extent to which CAI can complement and potentially enhance human designers' creative capabilities. By systematically evaluating the responses generated by CAI against the benchmarks set by these hypotheses, the study aims to provide empirical evidence on the viability of integrating AI into the ideation phase of the design process. This section details the outcomes of the pilot study, presenting a comprehensive analysis of the findings and their implications.

\subsubsection{6.2.1. Validation of Hypothesis-1}
\begin{figure}%[htbp]
\centerline{\includegraphics[width=0.9\columnwidth]{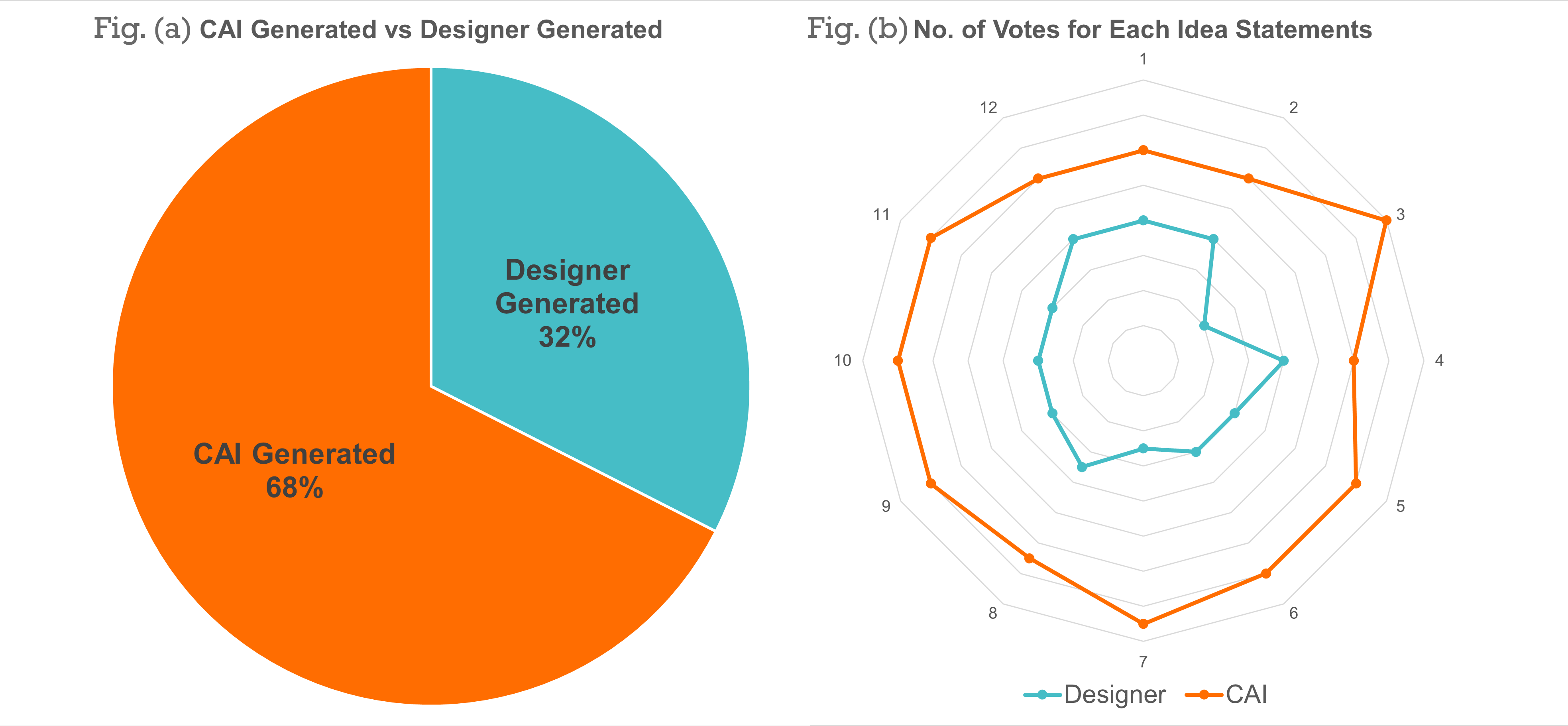}}
\caption{Pie Chart depicting the Average Voting for the Meaningfulness of the Ideas Generated by Designers and CAI}
\label{fig.meaningful_idea_chart}
\end{figure}

An expert-driven validation process was used to compare the responses generated by the CAI-based ideation tool with ideas generated by human designers. A questionnaire presented \emph{six pairs} of ideas wherein one idea in each pair was picked from Part A of the activity and the other from Part B, in random order. A panel of \emph{twenty} expert product designers gave their opinion on which statement they perceived to be a more meaningful idea in each pair. The outcome is illustrated in Figure~\ref{fig.meaningful_idea_chart}(a) and Figure~\ref{fig.meaningful_idea_chart}(b). It can be observed in Figure~\ref{fig.meaningful_idea_chart}(a) that 68\% of the experts found the ideas produced by the GPT more meaningful. Moreover, Figure~\ref{fig.meaningful_idea_chart}(b) shows that the votes for the GPT-generated statements consistently exceeded those for designer-generated ideas. Thus, \emph{Hypothesis H1: CAI produces meaningful ideas, is validated!}\newline

\subsubsection{6.2.2. Validation of Hypothesis-2}
\begin{figure}[t!]
\centerline{\includegraphics[width=1.0\columnwidth]{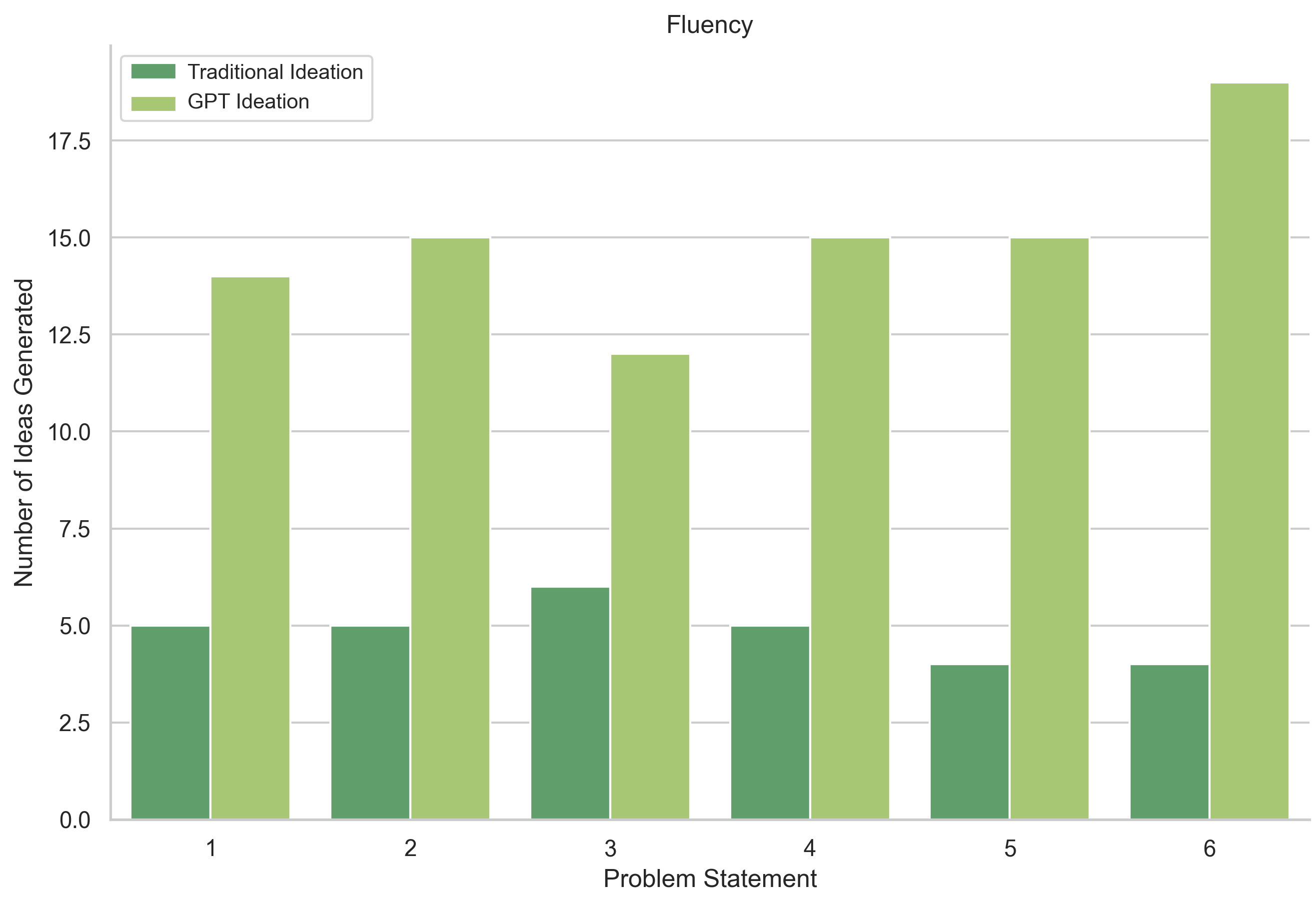}}
\caption{Bar plot of Fluency ($\Gamma$) - No. of Ideas Generated during Part A and Part B}
\label{fig.fluency}
\end{figure}

As indicated in Figure~\ref{fig.fluency}, the conventional vs CAI-aided ideation produced an average of 4.8 and 15 ideas in 20 minutes, respectively. Thus, fluency in Part B is nearly three times that of Part A. This notable increase in idea generation indicates a higher idea flow facilitated by GPT. Therefore, it supports the assertion that CAI can promote prolific ideation.

\begin{figure}[h!]
\centerline{\includegraphics[width=1.0\columnwidth]{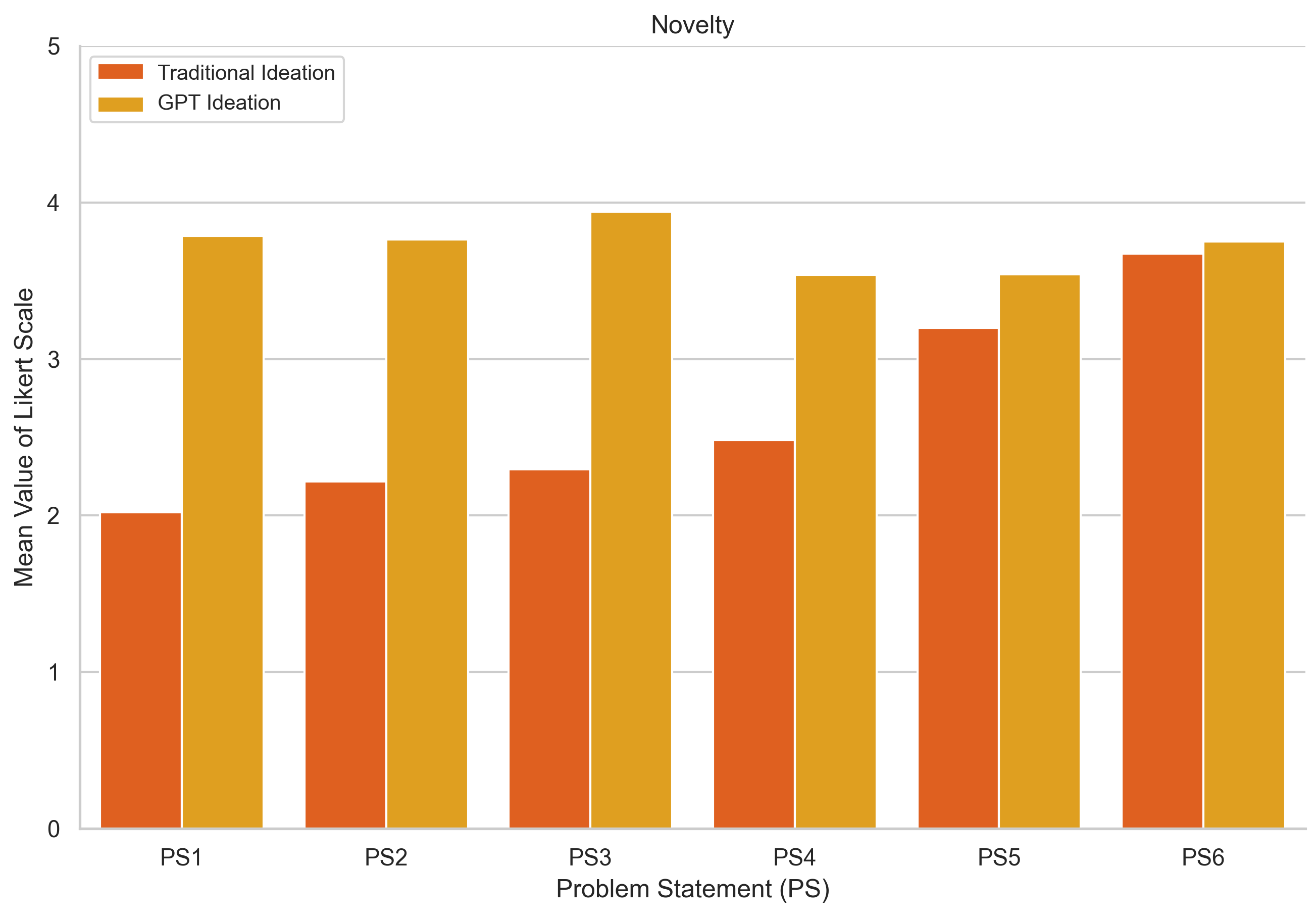}}
\caption{Bar Plot of Novelty ($\eta$) - Uniqueness of Ideas in Part A and Part B}
\label{fig.novelty}
\end{figure}

\begin{figure}[t!]
\centerline{\includegraphics[width=1.0\columnwidth]{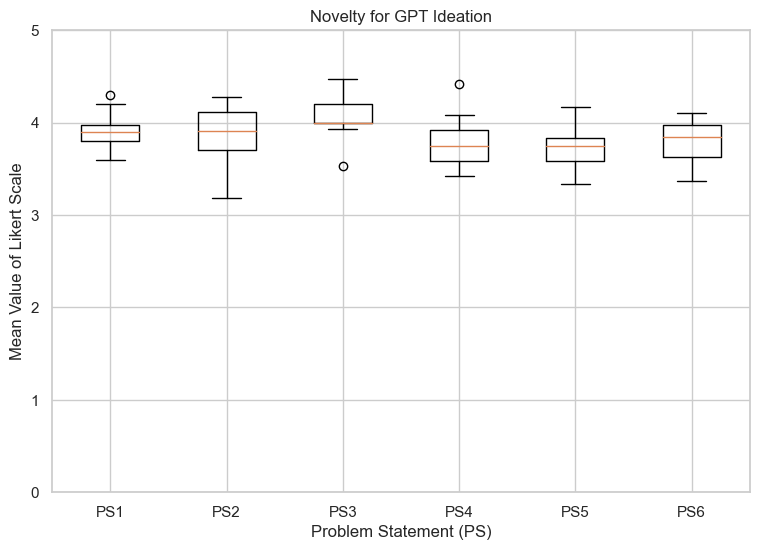}}
\caption{Box and Whisker Plot for Novelty in Part B (GPT Ideation)}
\label{fig.box_plot}
\end{figure}

The novelty ratings of individual ideas were evaluated on a 1 to 5 Likert Scale by 80 experts using an online questionnaire. The findings are depicted in Figure~\ref{fig.novelty}. It can be noted that the average rating for Part A is 2.5, while Part B is 3.86. Therefore, the ratings provided by experts significantly favoured CAI-enabled ideation over traditional techniques in terms of novelty. In addition to this, in the box plot analysis shown in Figure~\ref{fig.box_plot}, the ratings fall between 3.5 to 4.5, indicating a general consensus among experts that ideas generated by GPT have better novelty. Thus, \emph{Hypothesis H2: CAI promotes prolific novel ideas, is validated!}

\subsubsection{6.2.3. Validation of Hypothesis-3}
\begin{figure}[h!tbp]
\centerline{\includegraphics[width=1.0\columnwidth]{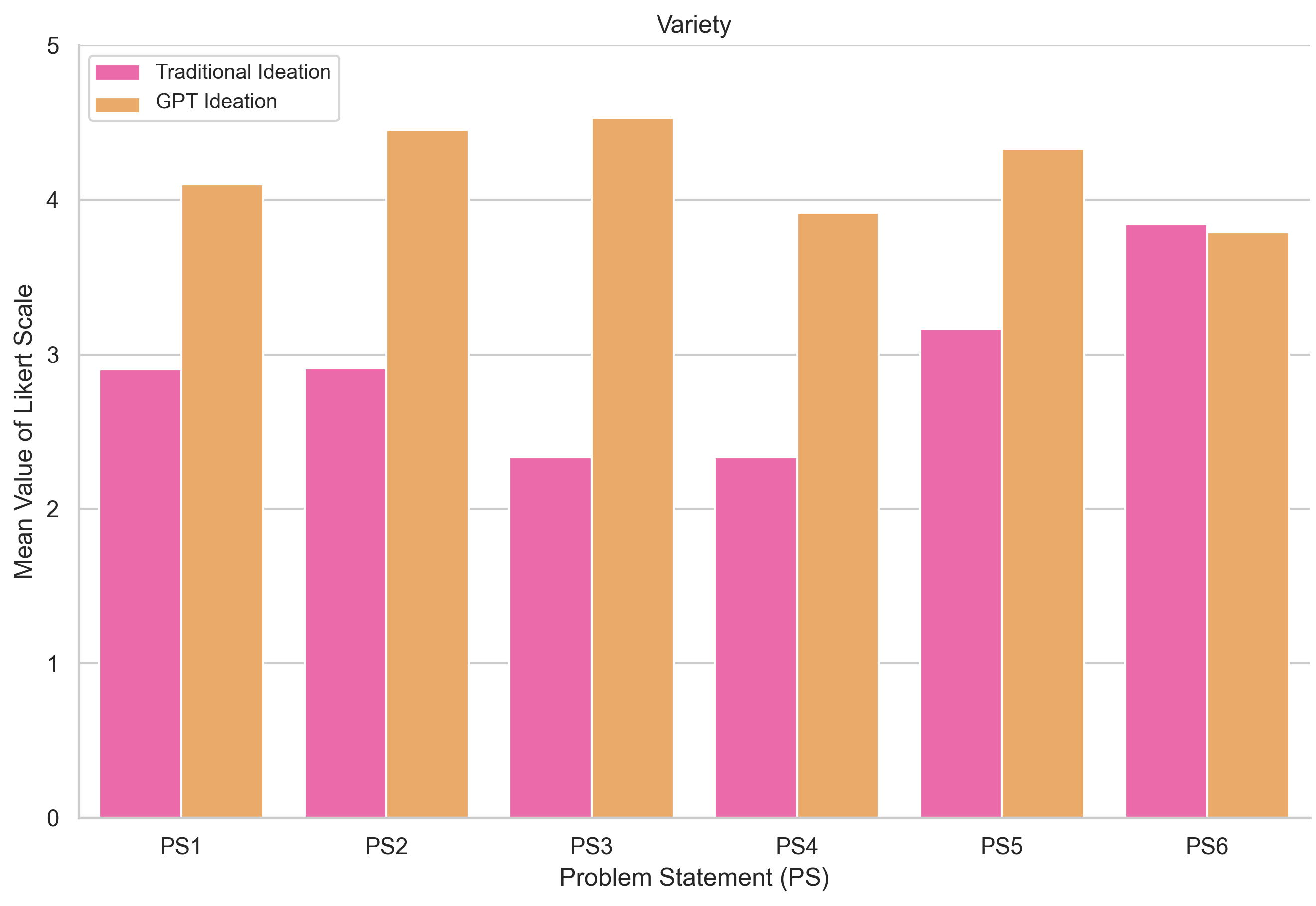}}
\caption{Bar Plot of Variety ($\upsilon$) - Diversity among Ideas in Part A and Part B}
\label{fig.variety}
\end{figure}

Variety or diversity of ideas is a key metric that reflects the effectiveness of an ideation tool for designers. The variety ratings of a set of twelve ideas randomly picked from a larger set were evaluated on a 1 to 5 Likert Scale by 80 experts using an online questionnaire. The protocol ensured that the sampling adequately covered the complete set of about 80 ideas. The findings are depicted in Figure~\ref{fig.variety}. The results show that the average variety rating for Part A is 2.9, while Part B is 4.2. Therefore, the ratings favoured CAI-enabled ideation over traditional methods. This considerable increase in variety demonstrates that GPT generates more diverse ideas. Thus, \emph{Hypothesis H3: CAI promotes diverse ideas, is validated!}

\subsection{6.3. Inferences and Discussion}
Every idea can be associated with some cognitive resource (wisdom - experience, knowledge - expertise, information - learning, data - observation, etc., as prescribed by the DIKW pyramid from information science) of the designer(~\cite{Saulais2023}). Table~\ref{tab.idea_dik} illustrates this association for some ideas from Table~\ref{tab.set_of_ideas}. This implicit connection between knowledge and idea demands that the designers can connect their cognitive resources to the problem at hand while proposing novel ideas. However, voluntary recalling of relevant knowledge is a complex yet unreliable cognitive process. On the other hand, a CAI system is inherently efficient in associating a query with the resources it has been trained with. Given its training using the vast knowledge base, GPT has quick access to in-depth knowledge than an individual designer in any particular domain of relevance. Thus, CAI-enabled ideas scored higher in novelty consideration.

\begin{table*}[h!]
    \centering
    \caption{Association of Idea with Knowledge}
    \label{tab.idea_dik}
    \rowcolors{2}{lightgray!80}{lightgray!40}
    \begin{tabular}{|m{1.5cm}|m{9.0cm}|m{6.0cm}|}
    %\begin{tabular*}{\columnwidth}{@{\extracolsep{\fill}}|l|c|r|}
    %\begin{tabularx}{\columnwidth}{|p{1.5cm}|p{1.5cm}|X|}
    \hline
       \centering
       \textbf{Problem Statements (PS)} & 
       \centering
       \textbf{Generated Ideas} & 
       \textbf{Associated Knowledge}
       \\
    \hline
        PS1 
        & 
        CI-1.2. Incentivized Waste Segregation: Implementing an incentive-based system where users receive rewards or benefits for proper waste segregation and disposal. This encourages active participation and responsibility.
        & 
        Rewards and Punishments Influence Human Behaviour. 
        \\
    \hline
        PS2 
        & 
        CI-2.2. Liquid Jet Spray Chamber: Developing footwear cleaning chambers equipped with liquid jet spray mechanisms that shoot cleaning solutions onto shoes. Users would enter the chamber, and the liquid jets thoroughly clean and disinfect their footwear.
        & 
        High-Pressure Liquid Jets help remove micro-particles from a surface.
        \\
    \hline
        PS3
        & 
        CI-3.2. Dishwashing Gloves: Developing dishwashing gloves with built-in scrubbing surfaces on the palms and fingers allows users to scrub dishes without needing separate scrubbing tools.
        & 
       Wearables improve tactile control and are ergonomically efficient in doing a task.
        \\
    \hline
        PS4
        & 
        CI-4.2. Queueing Cushion: Developing cushions designed specifically for queueing featuring ergonomic shapes and materials that provide comfortable seating options for individuals waiting in line with Inserts Straps
        &
        Soft foams expand four times the volume with which they can be stored and are adaptable and lightweight.
        \\
    \hline
        PS5 
        &
        CI-5.2. Automatic Gravity-Enabled Feeder: Creating an automatic bird feeder that uses gravity to dispense feed as needed reduces manual refilling efforts and ensures a consistent food supply.
        &
        Gravity-driven mechanisms are continuous and self-regulating, minimizing the need for manual intervention.
        \\
    \hline
        PS6
        & 
        CI-6.2. Electrostatic Dryer: Designing umbrellas with an electrostatic drying mechanism that is battery-activated and repels water from the canopy when activated, ensuring a dry umbrella before coming indoors.
        &
        Electrostatic forces can be harnessed to repel water molecules from surfaces, effectively creating a barrier that prevents water from clinging and facilitates rapid drying.
        \\
    \hline
    \end{tabular}
    %\end{tabular*}
    %\end{tabularx}

\end{table*}

Human designers often reach a saturation point where they cannot generate new ideas. This is referred to here as a mental block (one of the ideation bottlenecks). A broader knowledge base from \emph{multiple domains} is important in helping the designers break free from these ideation bottlenecks~(\cite{Mostert2007}).  GPT is trained in knowledge across many domains; its inherent stochastic nature ensures that different solutions are generated when queried repeatedly. A designer's ability to acquire and access knowledge is limited to a select few areas of expertise. Therefore, GPT, unlike humans, does not exhibit saturation and generates more diverse ideas.

\begin{figure}[h!tbp]
\centerline{\includegraphics[width=1.0\columnwidth]{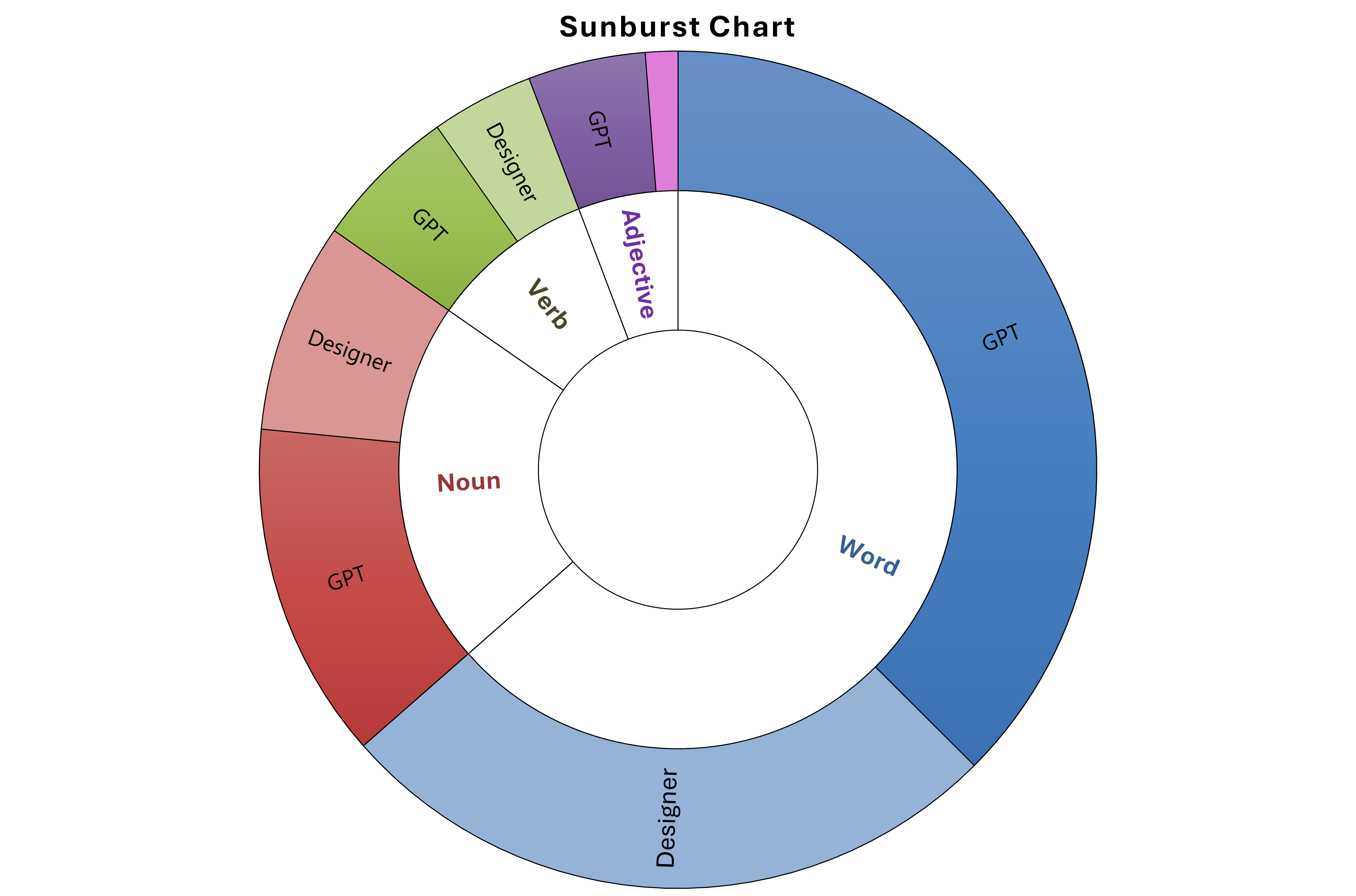}}
\caption{Linguistic Element Breakdown for the Ideas Generated by Designer and CAI}
\label{fig.sun_burst_linguistics}
\end{figure}

The linguistic analysis conducted on the ideas generated by designers and GPT, as shown in Figure~\ref{fig.sun_burst_linguistics}, reveals that GPT, on average, has a higher count of linguistic elements such as words, nouns, verbs, and adjectives. These elements serve as technical descriptors within a language, indicating that CAI-generated ideas possess greater detail than those conceived by human designers. Also,
Figure~\ref{fig.meaningful_idea_chart}(b) showed an interesting outcome: the average number of votes for each idea statement was higher for CAI than for the designer across all problem statements. A few expert designers were personally interviewed to understand the rationale behind their choices. They acknowledged that the level of detail contained in a CAI-generated idea was richer, providing them with more insights about the ideas. This depth of detail can be seen as an advantage in providing a rich context for each solution, but there is also a potential risk of cognitive overload. Designers could potentially become overwhelmed by the surplus of information. This would detract the idea, distract the designers from their objective, and diverge them from the original problem.

From the above discussion, it is understood that CAI-aided ideation is more beneficial in the ideation process than traditional methods (RQ1 is Answered). It helps the designers break free from the ideation bottlenecks that hold them back from potential solutions (RQ2 is Answered). By integrating CAI into their workflow, designers shift their primary role from generating ideas to curating them. This pivot allows designers to leverage their expertise in elaborating and refining the CAI's suggestions. This collaborative approach between machine creativity and human selectivity has shown the potential to enhance the efficiency of the ideation phase.

\section{7. Structuring the Interaction Style with CAI }
The pilot study has highlighted the capabilities of Conversational AI (CAI) in producing a rapid, detailed, novel and diverse collection of ideas. However, the efficacy of CAI is intricately linked to the specificity of the queries posed to it. Thus, to enhance the effectiveness of the CAI, there is a need for a consistent, structured format for the queries (prompts) given as input to the CAI. This structure should encapsulate the essential requirements of design, making it uniformly effective for any designer using our CAI system. On the other hand, it also presents a challenge, as the depth and volume of details generated by CAI can lead to cognitive overload during the selection phase, where designers must sift through extensive information to identify and select potential solutions. This poses a daunting task, given the huge number of ideas generated by the CAI. Thus, there is a need for a consistent, structured format for the responses generated as output by the CAI similar to that of the prompts. This ensures quick information retrieval and fast interpretation by the designers, reducing uncertainty in navigating the generated content. Novice designers struggle to \emph{formulate a clear problem statement from the user needs, generate novel ideas and create feasible concepts}. Thus a designer requires assistance in all these three tasks. Structured prompts and responses have been identified as a critical strategy in managing the flow of information, ensuring that responses from LLMs are both relevant and succinct (\cite{Lynch2023}). Therefore, a structured input and output balances the richness of information generated by CAI with the cognitive capacity of novice designers. Thus, we propose a novel structure for articulating problem statements from unstructured user needs, generating structured ideas from structured input prompts and synthesizing structured concepts from structured ideas within the CAI framework.

\subsection{7.1. Structuring the Problem Statement}
\begin{figure}[htbp]
%\centerline{\includegraphics[width=1.0\columnwidth]{Images/Problem Structure v2.png}}
\centerline{\includegraphics[width=1.0\columnwidth]{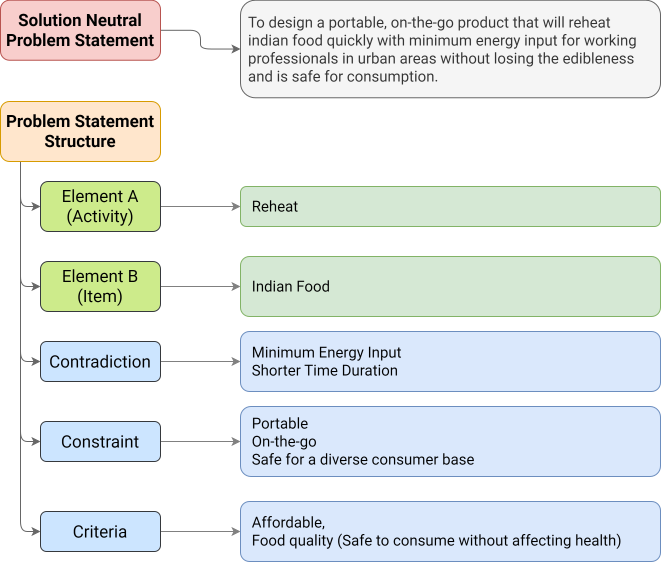}}
\caption{Illustrative Example of a Problem Statement Structure}
\label{fig.problem_structure}
\end{figure}

Defining and understanding a problem's nature is important for producing novel solutions in product design~(\cite{CHDorst2003}).  The use of Solution Neutral Problem Statement (SNPS) is a standard practice for this purpose. However, designers often find it difficult to state the problem in this format. Hence, we are proposing a structure that will be used by the CAI to systematically formulate the problem statement by pinpointing the challenges that make the problem difficult, given the user needs as input by the designer. Ideation involves propositions by which these challenges could be overcome. The proposed structure for a problem statement is encapsulated in the "\textit{AI3C: Activity-Item-Contradiction-Constraint-Criteria}" model (An Example of the problem statement structure is shown in Figure~\ref{fig.problem_structure}).

The proposed problem structure is based on the interaction of two critical elements:
\begin{enumerate}
    \item Element A (Activity): This represents the action or series of actions applied within the problem space.
    \item Element B (Item): This is the target or subject upon which the activity is performed, leading to a desired state change.
\end{enumerate}

Under ideal conditions, where knowledge and resources are available, the interaction between Element A and Element B should not pose any challenge, and state change should be readily achievable. However, in practical scenarios, the interplay of the Contradiction, Constraint, and Criteria complicates this interaction, giving rise to a problem, thereby making the core issue explicit.

Contradiction arises when there is a direct conflict between the desired state change and the actual relationship between Element A and Element B. 

Constraints are the bounding conditions within which the problem must be solved. These may include technical constraints, such as the maximum weight a material can support; economic constraints, such as budget limits; or regulatory constraints, such as safety and environmental regulations.

Criteria represent the benchmarks for evaluating the success of a solution. They are the qualitative and quantitative goals the design must achieve to succeed. These could include performance criteria like speed or efficiency, usability criteria such as user-friendliness, or environmental impact criteria such as carbon footprint.

\subsection{7.2. Structuring the Input Prompt Style for Idea Generation}
\label{sec:structuring_input_prompt_style}
The response generated by CAI engines depends upon the exact phrase used as input for a given query. Therefore, it is important to formulate the right style for the phrases to invoke more useful responses in the design context. Ideation is a common cognitive activity during the conceptual design phase, which \emph{generates propositions} by which challenges could be overcome for the problem. The cognitive process of problem-solving involves the stages of searching, learning, synthesis, analysis, and inference~(\cite{b25, b26, b27}). During recursive ideation, the corresponding stages are referred to as exploration (searching), inspiration (learning), generation (synthesis), elaboration (analysis), and evaluation (inference). We map the different types of prompts available in the domain of prompt engineering~(\cite{b28, b29}) in CAI to be given as input that facilitates each of these ideation stages. A prompt is an assertive sentence used to solicit a response from CAI. Each type of prompt is characterized by a context and a query. A context is defined by the nature of the fields contained in it. The essential fields for each context and the corresponding structure of the prompt are presented in Table~\ref{tab.stages_in_ideation} with illustrative examples for each stage of ideation.

The key to harnessing the ideation potential of CAI lies in the structured formulation of input prompts and output responses. By carefully crafting prompts, designers can guide CAI in generating structured responses pertinent to the problem. In this way, LLMs are not bottlenecked by the designer's limited knowledge or the stochastic nature of human cognitive recall. Providing a structure for inputs guides the CAI in understanding the context and constraints of the problem space, ensuring that the generated ideas are relevant and focused. Without such a structure, the CAI produces responses that, while potentially creative and diverse, lack specificity to the task at hand and overwhelm the user with an excess of detailed procedural content, as seen from the results of the pilot study (Table ~\ref{tab.set_of_ideas}). Structured prompts help steer the AI towards the generation of 'what to do' ideas rather than 'how to do it' processes, thus aiding in the conceptual phase of solution development.

\begin{table*}[h!]
    \centering
    \caption{Stages in Ideation with the Corresponding Prompts, Essential Fields in each prompt with an Example}
    \label{tab.stages_in_ideation}
    \rowcolors{2}{lightgray!80}{lightgray!40}
    \begin{tabular}{|m{1.5cm}|m{1.5cm}|m{3.0cm}|m{9.0cm}|}
    %\begin{tabular*}{\columnwidth}{@{\extracolsep{\fill}}|l|c|r|}
    %\begin{tabularx}{\columnwidth}{|p{1.5cm}|p{1.5cm}|X|}
    \hline
       \centering
       \textbf{Stages in Ideation} & 
       \centering
       \textbf{Name of the Prompt} & 
       \centering
       \textbf{Essential Context Fields} &
       
       \textbf{Illustrative Prompt for CAI}\\
    \hline
        Exploration Stage 
        & 
        Role Prompts
        & 
        1. Profession \newline
        2. Domain \newline
        3. Considerations \newline
        4. Priorities \newline
        5. Questions
        &
        \textbf{Context}: Assume the role of a [\emph{environmental scientist}] with expertise in [\emph{water quality and purification technologies}]. I need your insights on [\emph{purifying water from natural sources}] using methods that are [\emph{environmentally sustainable and effective}]. Answer the following [question(s)]:\newline
        \textbf{Query}:\newline
        \emph{What technologies are used in current systems for water purification?}
        \emph{Considering the weight and durability, what materials do you recommend?}
        \\
    \hline
        Inspiration Stage 
        & 
        Shot Prompts
        & 
        1. Inspirations \newline
        2. Analogous Situations \newline
        3. Domains \newline
        4. Mechanism 
        &
        \textbf{Context}: Draw inspiration and analogous situations, processes, and solutions from [\emph{nature and biomimicry}] focusing on [\emph{natural filtration and purification mechanisms}]. \newline
        \textbf{Query}:\newline
        Provide examples of the same that could inspire potential solutions for my design.\\
    \hline
        Generation Stage 
        & 
        Open-Ended Prompts
        & 
        1. Action \newline
        2. Problem \newline
        3. Included Domains \newline
        4. Excluded Domains
        &
        \textbf{Context}: Imagine a novel approach to [\emph{purifying water}] that addresses [\emph{the removal of a wide range of contaminants from various water sources encountered in the wilderness}]. Consider methods, technologies, and/or processes that combine elements from [\emph{biomimicry, material science, and renewable energy}] and the ones that have not been traditionally associated with [\emph{water purification}].\newline 
        \textbf{Query}:\newline
        What might such a solution look like, and what innovative features could it include? Describe how these features could address the problem uniquely and improve it. Feel free to think outside the box and propose ideas that might seem unconventional or futuristic.\\
    \hline
        Elaboration Stage 
        & 
        Leading Prompts
        &
        1. Idea \newline
        2. Goal \newline
        3. Aspects \newline
        4. Add-Ons
        &
        \textbf{Context}: Consider the initial idea of [\emph{solar-powered sterilization unit}]. Let's delve deeper into this idea. How could we enhance this idea better to achieve [\emph{greater efficiency in removing a wider range of contaminants while maintaining portability and durability}]? Consider [\emph{usability in diverse environmental conditions, energy efficiency, and the use of sustainable materials}]. \newline
        \textbf{Query}:\newline
        Can we integrate a [\emph{biomimetic filtration system}] into this idea? \newline
        Provide a detailed description of this enhanced idea.\\
    \hline
        Evaluation Stage 
        &
        Option Prompts
        &
        1. Idea 1\newline
        2. Idea 2\newline
        3. Constraints \newline
        4. Requirements
        &
        \textbf{Context}: Consider the shortlisted ideas: [\emph{Idea 1: A solar-powered UV water purification device}], [\emph{Idea 2: A manual, pump-operated filter system using biodegradable filters}]. Given the [\emph{limited access to power sources, the need for lightweight and compact solutions, and environmental sustainability}]. \newline
        \textbf{Query}:\newline
        Compare the aforementioned ideas in terms of their [\emph{effectiveness in contaminant removal, ease of use, sustainability, and portability}]. \newline
        Which idea(s) would be more effective? Could these ideas be combined?  \newline
        Provide critical feedback on each idea, highlighting strengths, weaknesses, opportunities, and threats.\\
    \hline
    \end{tabular}
    %\end{tabular*}
    %\end{tabularx}
    
\end{table*}

\subsubsection{7.2.1. Exploration Stage: Role Prompts}
During the exploration stage, the designer looks for existing solutions externally in patents, academic research, markets, etc. and internally in their experience and memory that wholly or partially address the problem at hand. The efficacy of the exploration depends on the \emph{expertise} of the person and the \emph{richness} of the resource. The richness in the present context will be determined by the specific CAI system adopted. Similarly, in CAI, a \emph{Role prompt} is designed to position CAI as a domain expert, biasing it to source solutions and knowledge from specific fields. For example, if prompted to play the role of a designer, its response to subsequent queries will entail suggestions grounded in design principles, creative problem-solving techniques, industry-specific knowledge, etc. 

% \noindent A generalized structure of a role prompt is given below:
% Assume the role of a [\emph{specific profession}] with expertise in [\emph{specific domain}] and insights on [\emph{relevant challenges/considerations}]. From your standpoint on [\emph{specific priorities}], Answer the following [\emph{guided question(s)}].

\subsubsection{7.2.2. Inspiration Stage: Shot Prompts}
With an understanding of the gap in the existing solutions, the designer advances to the inspiration stage, seeking stimuli from various domains, such as nature, scientific disciplines, and the external environment. Similarly, in CAI, \emph{Shot Prompt} is employed to direct CAI to provide concentrated facts (in short phrases) from diverse domains. A shot prompt is \emph{crafted} to extend beyond the immediate problem domain, encouraging the CAI to draw parallels and identify analogous situations or solutions in seemingly unrelated fields. It stimulates lateral thinking by cross-pollinating ideas from different domains, thereby fostering novel solutions.

% \noindent A generalized structure of a shot prompt is given below:
%Draw inspiration and analogous situations, processes, and solutions from [\emph{specific domain(s)}] focusing on [\emph{specific mechanism}]. Provide examples of the same that could inspire potential solutions for my design.

\subsubsection{7.2.3. Generation Stage: Open-Ended Prompts}
In the generation stage, the designer leverages their creatively stimulated mind to conceive and formulate new ideas for the given problem. Similarly, in CAI, \emph{Open-Ended prompt} is intentionally vague and broad, allowing for a wide range of creative responses. Open-ended prompts encourage CAIs to respond divergently, proposing novel ideas without constraints. This stage is crucial for brainstorming and expanding the horizon of potential solutions, as the CAI generates creative outputs that can be further refined.

% \noindent A generalized structure of an open-ended prompt is given below:
% Imagine a novel approach to [\emph{specific action}] that addresses [\emph{specific problem}]. Consider methods, technologies and/or processes that combine elements from [\emph{different domains}] and the ones that have not been traditionally associated with [\emph{specific domain}]. What might such a solution look like, and what innovative features could it include? Describe how these features could address the problem uniquely and improve. [\emph{Feel free to think outside the box and propose ideas that might seem unconventional or futuristic.}]

\subsubsection{7.2.4. Elaboration Stage: Leading Prompts}
Following the generation of a basic idea, the elaboration stage involves a deeper contemplation and refinement of an idea to better align it with the potential solution. Similarly, CAI, \emph{Leading Prompt} help the designer to guide CAI with specific examples or scenarios, prompting it to expand further and detail the idea. These prompts are targeted, asking the CAI to elaborate on particular aspects of the idea, enriching the idea and adding depth to the proposed solution.

% \noindent A generalized structure of a leading prompt is given below:
% Consider the initial idea of [\emph{specify the idea}] aimed at addressing the [\emph{specific problem}]. Let's delve deeper into this idea. How could we enhance this idea better to achieve [\emph{specific goal}]? Consider [\emph{specific aspects}]. Can we integrate a [\emph{specific feature/technology/process/another idea}] from [\emph{specific domain}] to this existing idea? Provide a detailed description of this enhanced idea. Include considerations for [\emph{specific implementation/interaction/impact}].

\subsubsection{7.2.5. Evaluation Stage: Option Prompts}
The final stage, evaluation, occurs when the designer has a set of promising ideas and seeks to either select the most viable ones or amalgamate them to forge new concepts. Similarly, in CAI, \emph{Option Prompt} is used by the designer in presenting CAI with a set of shortlisted ideas and instructing it to assess them to evaluate the ideas. These evaluative and comparative prompts enable the CAI to provide critical feedback or combine elements from different ideas to create superior solutions.

% \noindent A generalized structure of an option prompt is given below:
% Consider the shortlisted ideas: [\emph{idea 1}], [\emph{idea 2}], [\emph{idea 3 (if needed)}], each aiming to address [\emph{specific problem}].Given the context of [\emph{specify conditions/constraints}] and considering the [\emph{specific need(s)}] for [\emph{target user}], Compare the aforementioned ideas in terms of their [\emph{specific requirement(s)}]. Which idea(s) would be more effective? Could these ideas be combined? Provide critical feedback on each idea, highlighting strengths, weaknesses, opportunities and threats.

A problem statement was taken as an example to illustrate the corresponding prompts and their essential fields. The problem statement is to create an eco-friendly portable water purification device for hikers. This device should be lightweight, durable, and capable of removing contaminants from various water sources encountered in the wilderness. Table~\ref{tab.stages_in_ideation} displays an example of each prompt in the defined structure that designers can use for respective stages of ideation to converse with CAI.\newline

%----------------------------------------------------------------------
\begin{figure}[t!]
%\centerline{\includegraphics[width=1.0\columnwidth]{Images/Idea Structure v2.png}}
\centerline{\includegraphics[width=0.9\columnwidth]{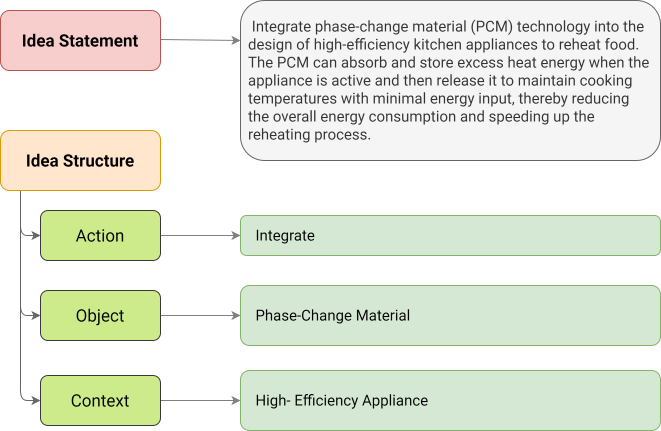}}
\caption{Illustrative Example of an Idea Structure}
\label{fig.idea_structure}
\end{figure}

\subsection{7.3. Structuring the Output Response Style For Generated Ideas and Concepts}
CAI systems have the propensity to deliver verbose content, leading to \emph{information overload}~(\cite{zhang-Yue-2023, Ye-Hongbin-2023, Huang-Lei-Yu-2023, Xu-Ziwei-2024}). This impedes the designer's ability to rapidly \emph{assimilate and act} upon the information provided. Thus, parallel to establishing a standardized format for input prompts, as seen in the earlier section, a structured format is deemed necessary for the responses given by CAI. This would ensure that the output from CAI is uniform, concise, coherent, consistent and tailored to the designer's specific needs.

\begin{figure}[t!]
% \centerline{\includesvg[width=0.9\columnwidth]{Images/Concept Structure v3.svg}}
\centerline{\includegraphics[width=0.9\columnwidth]{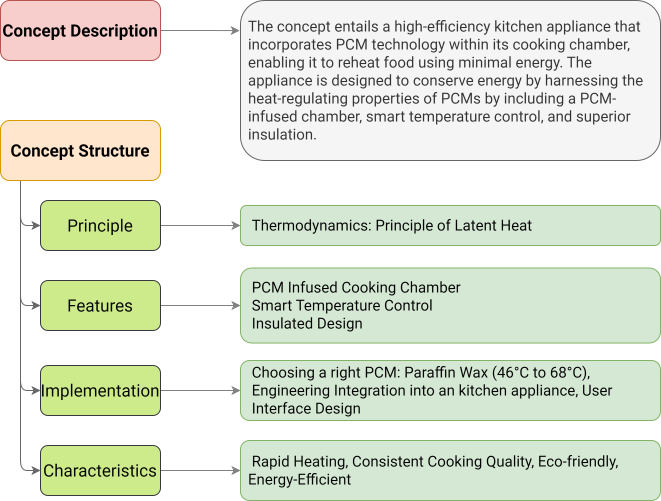}}
\caption{Illustrative Example of a Concept Structure}
\label{fig.concept_structure}
\end{figure}

\subsubsection{7.3.1. Structuring the Response for Idea}
Historically, the articulation of ideas has predominantly taken the form of natural language sentences, which inherently consist of a subject and a predicate. These sentences serve as the medium through which designers express the connections they draw between a given problem and their reservoir of knowledge. The composition of these sentences often reveals the underlying structure of the ideation process, elucidating the "\emph{what}" component of the problem-solution space, i.e., what action (verb) applied to what object (noun) could potentially constitute a viable approach to addressing the challenge at hand.

This linguistic representation of ideas is a deeply rooted practice within the design domain, reflecting the natural tendency of designers to think and communicate in structured language patterns~(\cite{b11}). The prevalence of this practice provides a foundation upon which we establish a formalized structure for the presentation of ideas. The proposed structure for an idea is encapsulated in the  "\textit{AOC: Action-Object-Context}" model (An Example of the idea structure for a problem is shown in Figure~\ref{fig.idea_structure}). 

\emph{Action} is a verb that represents the transformative step or approach proposed to tackle the problem. It is the dynamic aspect of the idea, indicating how the designer envisions altering the current undesirable state. 

\emph{Object} is a noun that specifies the item or entity the action targets or involves. It is the focal point of change, the recipient of the action's effects. 

\emph{Context} provides the setting or environment in which the idea is situated, offering additional dimensions and considerations that may impact the idea's implementation and efficacy.

\begin{figure*}[t!]
\centerline{\includegraphics[width=1.0\textwidth]{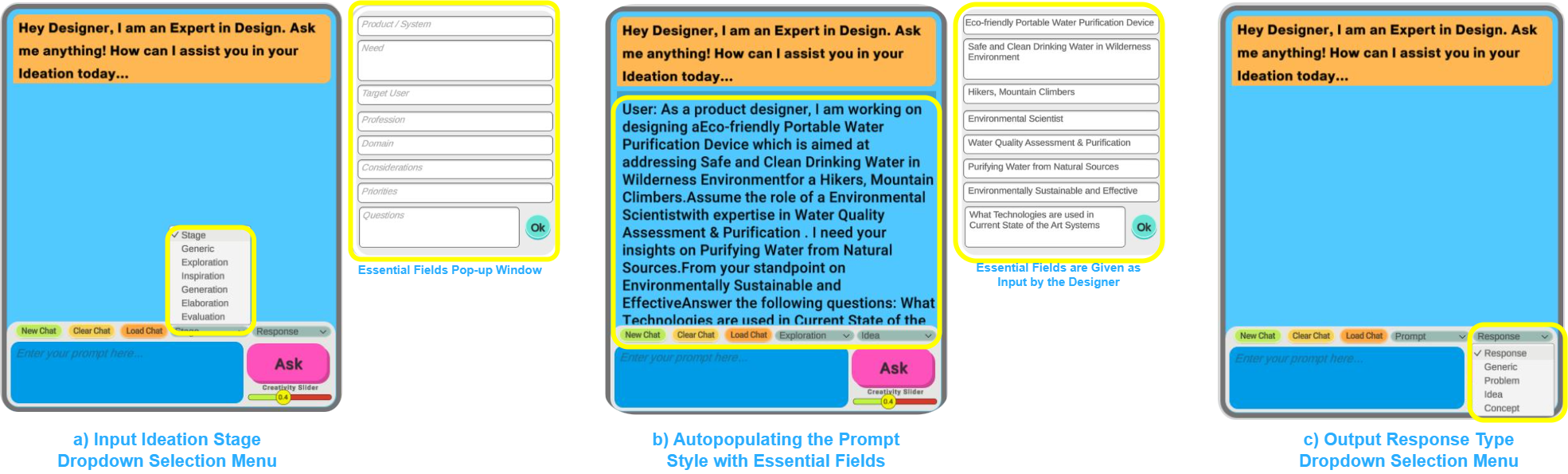}}
\caption{Interaction Style Selection in CAI Interface}
\label{fig.interaction_style}
\end{figure*}

\subsubsection{7.3.2. Structuring the Response for Concept}
The transition from ideation to the concrete development of a concept in the design process requires a meticulous and structured approach to ensure that the nascent ideas are transformed into viable solutions. In the design domain, a concept is a proposal that outlines the practicality and the technicalities of "\emph{how}" an idea can be realized. It is a blueprint that follows the idea and must be grounded in scientific principles to ensure feasibility. The proposed structure for a concept is encapsulated in the  "\textit{PFIC: Principles-Features-Implementation-Characteristics}" model (An Example of the concept structure for a problem is shown in Figure~\ref{fig.concept_structure}.

\emph{Principles} refer to the scientific laws, theories or techniques that underpin the concept and guarantee its feasibility. These principles are the bedrock upon which the concept stands, providing the rationale and validation for why and how the concept can work in real-world scenarios.

\emph{Features} detail the various components or attributes of the concept.  These are the tangible elements that differentiate one concept from another, providing a clear picture of the concept's design and functionality.

\emph{Implementation} outlines the method or process by which the concept will be realized. It bridges the gap between theory and practice, ensuring that the concept can be operationalized and that the transition from paper to prototype is feasible.

\emph{Characteristics} define the qualities or behaviours of the concept, often described by adjectives or interjections. These descriptors will define the concept's performance, usability, and overall impact. Characteristics provide insight into the concept's interaction with its environment and end-users.

In our GPT-based Ideation Interface discussed in Section~\nameref{sec:ideation_interface}, two separate drop-down menu options are provided for selecting the ideation stage that corresponds to the input prompt type and output response type as shown in Figure~\ref{fig.interaction_style}. By choosing the ideation stage, a pop-up window requires the designers to fill in the essential fields. The input message auto-populates with the corresponding structure of the prompt with essential fields given by the designer. It is the decision of the designer to choose the right stage depending on their requirement. Each prompt type initiates a \emph{new CAI conversation} whose context is preset based on the user input. The choice of output selection provokes the CAI to provide the corresponding response in the structure defined within the CAI system.

\begin{figure}[t!]
% \centerline{\includesvg[width=1.0\columnwidth]{Images/Various Ideations v5.svg}}
\centerline{\includegraphics[width=1.0\columnwidth]{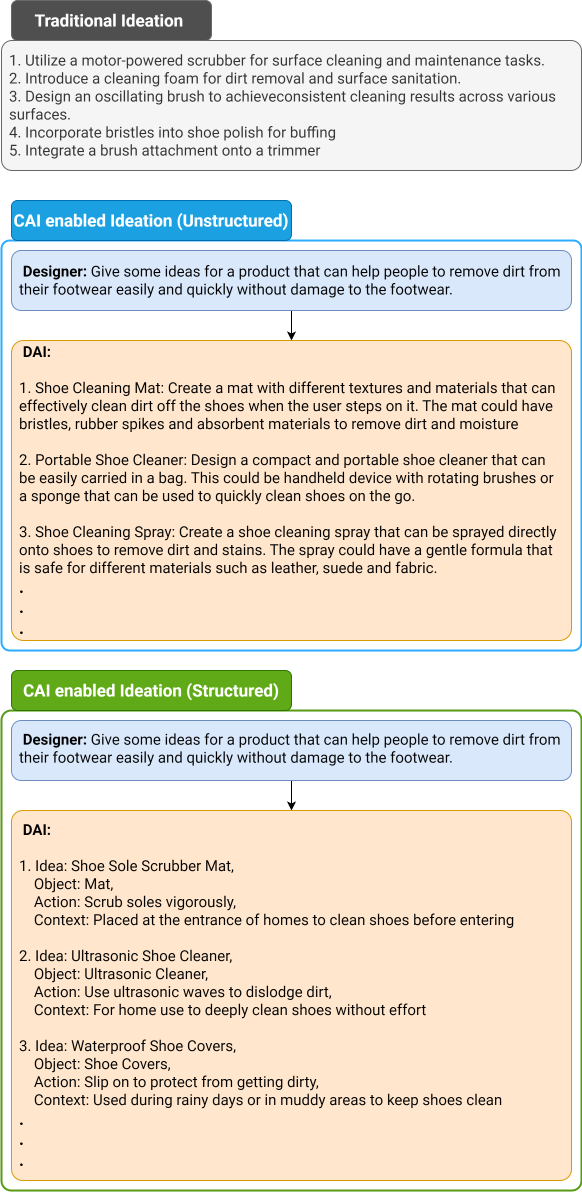}}
\caption{Outcome of Traditional Ideation vs CAI enabled Ideation (Unstructured and Structured)}
\label{fig.various_ideations}
\end{figure}

%-----------------------------------------------------------

\subsection{7.4. Analysis of Unstructured vs Structured CAI Ideation}

A comparative analysis was conducted to evaluate the effectiveness of the traditional brainstorming methods and the unstructured and structured CAI-based ideation processes. For this analysis, a problem statement was taken that addressed people's difficulty in cleaning footwear quickly and easily without causing damage to it. The study aimed to show the differences in the relevance of the ideas generated through these distinct ideation techniques. The ideas generated through these methods are shown in Figure~\ref{fig.various_ideations}.

%Traditional Ideation via Brainstorming
Utilizing the traditional brainstorming technique, a group of novice designers engaged in a session to generate solutions for the stated problem. The ideas produced, while diverse, largely reflected solutions that were analogous to those used for cleaning other items, such as household or automotive cleaning tools. This outcome may be attributed to the 'Lack of Experience' bottleneck, where participants reverted to familiar concepts rather than innovating new ones specifically tailored to footwear. In our view, 'Design Fixation' also played a role, limiting the exploration beyond the initial ideas, which was evident from the repetitive use of brushes/bristles in the ideas generated.

%Unstructured CAI-Based Ideation
In contrast, the unstructured CAI-based ideation process, driven by a conversational AI system, yielded a detailed array of ideas. The CAI, leveraging its access to extensive data repositories, produced solutions that were thorough in their explanation of the 'how to' aspect of implementation. However, the wealth of information provided by the CAI system in this unstructured format led to a saturation of procedural details, overshadowing the ideation's core 'what to do' aspect. This discrepancy highlighted the need for a more focused ideation approach that could distil the essence of the solutions without overwhelming the designers with excessive implementation specifics.

%**Structured CAI-Based Ideation**
A structured approach was introduced to address the challenges observed in the unstructured CAI-based ideation. This involved refining the CAI's prompts to concentrate more on generating ideas that defined 'what' the solution should be rather than 'how' it should be implemented. The structured CAI-based ideation process demonstrated a significant improvement, as evidenced by the attached Figure~\ref{fig.various_ideations}. The CAI-generated ideas became more precise and centred around the conceptualization of potential solutions, offering crisp and clear solutions that were easy to grasp.

\subsection{ 7.5. Summary}
The results from the structured CAI-based ideation show the importance of guiding the CAI to focus on generating novel ideas through a structured format. The 'Designer as a Curator' model becomes particularly relevant here, where the designer's role evolves to evaluate and refine the ideas proposed by the CAI, harnessing the computational system's capabilities while applying human intuition and expertise. The results reveal that while traditional brainstorming can yield a breadth of ideas, these may be limited by cognitive bottlenecks such as experience and fixation. Unstructured CAI-based ideation can overcome these bottlenecks by providing a vast array of detailed solutions. Structured CAI-based ideation is demonstrated as a more effective approach, enabling the generation of targeted, novel ideas that are more likely to result in practical and innovative solutions for the specific challenge.

%-----------------------------------------------------

\section{8. Discussion}
Ideation culminates with idea selection, when designers evaluate each idea based on certain requirements.  The process is complex and tedious as it involves comparing the ideas against each other. The CAI system presented above is capable of generating many novel and diverse ideas quickly. This abundance makes the shortlisting process more challenging and resource-intensive, potentially leading to decision fatigue with possible inconsistencies. The authors felt the need for an automated system to assist designers in efficiently selecting the most promising ideas from the generated pool. The progress made in this regard is beyond the scope of this paper.

A structured format for input queries and output responses provides convenience to the designer and easy assimilation of information by the CAI system to generate appropriate and relevant responses. This paper proposes one way to structure the prompts and responses, showcasing a few examples based on the natural way of communication familiar to the designers. The assessment of the effectiveness of this approach to be measured based on expert evaluations is still pending. The diverse methods of prompt structuring are also being explored.

%-------------------------------------------------------
\section{9. Conclusion}
\label{sec:conclusion}
This paper presented a conversational AI-enabled active ideation paradigm to enhance the ideation process, particularly for novice designers. By leveraging a large language model (LLM) such as GPT, the proposed scheme alleviates the cognitive bottlenecks experienced by designers during conventional ideation and facilitates a dynamic, interactive, and prolific ideation environment. It was shown that CAI significantly enhances the ideation process by promoting active, context-sensitive, and responsive interactions. The metrics of fluency and novelty in idea generation provided compelling evidence of this enhancement. Thus, the statement of numerous ideas for the problem is generated automatically; the designer's role is now to curate and select potential ideas. This makes the ideation process cognitively less burdening, leading to promising outcomes and making it accessible for less experienced designers. The proposed prompt engineering approach with a structured format for formulating problem statements, generating ideas, and creating concepts could not be found elsewhere in the literature by the authors. This approach ensures essential details are systematically provided and generated outputs are easily interpreted. This begets the CAI responses to be more relevant with a controlled level of diversity in the generated ideas. The large number of ideas generated by the CAI model poses a significant challenge for expert judicious assessment. The results presented were obtained by distributing random subsets of a manageable number of about 12 ideas to a large group of 80 experts. This arrangement, though practical, is not ideal and convenient. The authors are currently developing automated evaluation strategies to analyze the idea landscape effectively.

%%==========================================================================%
%\begin{acknowledgement}
%Insert the Acknowledgment text here.
%\end{acknowledgement}

%%==========================================================================%
%\paragraph{Funding Statement}

%\paragraph{Competing Interests}

%A statement about any financial, professional, contractual or personal relationships or situations that could be perceived to impact the presentation of the work --- or `None' if none exists.

%\paragraph{Data Availability Statement}

%A statement about how to access data, code and other materials allowing users to understand, verify and replicate findings --- e.g. Replication data and code can be found in Harvard Dataverse: \verb+\url{https://doi.org/link}+.

%%==========================================================================%

%\endnote in some journals will behave like \footnote, and \printendnotes will not output anything. 
%\printendnotes

%\input{2.bibliography}
\printbibliography

%--------------------------------------------------------------------------------------

%\appendix

%\section{Example Appendix Section}

%Lorem ipsum dolor sit amet

%%==========================================================================%
\end{document}